\documentclass[a4paper,11pt]{article}
\pdfoutput=1

\usepackage{jcappub}
\usepackage[T1]{fontenc}
\usepackage{subcaption}
\usepackage{mathtools}
\usepackage{physics}
\usepackage{amsmath}
\usepackage[normalem]{ulem}
\usepackage[nottoc]{tocbibind}

\title{Testing the assumptions of the Effective Field Theory of Large-Scale Structure}
\author[a]{Mandar Karandikar}
\author[a]{Cristiano Porciani}
\author[b,c]{Oliver Hahn}

\affiliation[a]{Argelander-Institut f{\"u}r Astronomie, Auf dem H{\"u}gel 71, D-53121 Bonn, Germany}
\affiliation[b]{Department of Astrophysics, University of Vienna, Türkenschanzstraße 17, 1180 Vienna, Austria}
\affiliation[c]{Department of Mathematics, University of Vienna, Oskar-Morgenstern-Platz 1, 1090 Vienna, Austria}

\emailAdd{mandar@astro.uni-bonn.de}
\emailAdd{porciani@astro.uni-bonn.de}
\emailAdd{oliver.hahn@univie.ac.at}

\abstract{The Effective Field Theory of Large-Scale Structure (EFTofLSS) attempts to amend some of the shortcomings of the traditional perturbative methods used in cosmology. It models the evolution of long-wavelength perturbations above a cutoff scale without the need for a detailed description of the short-wavelength ones. Short-scale physics is encoded in the coefficients of a series of operators composed of the long-wavelength fields, and ordered in a systematic expansion. As applied in the literature, the EFTofLSS corrects a summary statistic (such as the power spectrum) calculated from standard perturbation theory by matching it to $N$-body simulations or observations. This `bottom-up' construction is remarkably successful in extending the range of validity of perturbation theory. In this work, we compare this framework to a `top-down' approach, which estimates the EFT coefficients from the stress tensor of an $N$-body simulation, and propagates the corrections to the summary statistic. We consider simple initial conditions, viz. two sinusoidal, plane-parallel density perturbations with substantially different frequencies and amplitudes. We find that the leading EFT correction to the power spectrum in the top-down model is in excellent agreement with that inferred from the bottom-up approach which, by construction, provides an exact match to the numerical data.  This result is robust to changes in the wavelength separation between the two linear perturbations. However, in our setup, the leading EFT coefficient does not always grow linearly with the cosmic expansion factor as assumed in the literature based on perturbative considerations. Instead, it decreases after orbit crossing takes place.}

\begin{document} 
\maketitle
\flushbottom

\section{Introduction}
On scales smaller than the Hubble radius, dark matter dynamics is governed by the Vlasov-Poisson (VP) system of equations. In order to study the formation and evolution of the large-scale structure of the Universe, two approaches are widely used to solve the VP system: $N$-body simulations and perturbative techniques. The former approximate the phase-space distribution by a set of macroparticles that are gravitationally evolved \cite[e.g.][]{davis1985, bertschinger1998}. In the latter, we take velocity moments of the Vlasov equation, and solve the resulting equations perturbatively in the Eulerian or Lagrangian frame \cite[e.g.][]{bernardeau2002, bouchet1994perturbative}. To make this viable, the hierarchy of Vlasov moments is truncated at second order, by assuming the vanishing of the second cumulant -- the stress tensor. This truncation is justified in the cold limit before the crossing of distinct streams of matter (orbit crossing). The physical picture is that of a pressureless, perfect fluid.

Perturbation theory has been successful in advancing our understanding of structure formation on large scales \cite[e.g.][]{jain1994second, scoccimarro1997cosmological, crocce2006renormalized, taruya2009nonlinear, carlson2009critical, matsubara2008resumming}. However, it has two important shortcomings. First, non-linear gravitational clustering breaks the assumption of negligible velocity dispersion on successively larger scales, and second, it treats non-linear scales as perturbative. These are non-trivial problems, as at the instant of orbit crossing, all higher-order Vlasov cumulants are simultaneously generated, giving rise to a closure problem for the hierarchy. If we are interested in the overdensity and velocity fields, we can close the system of equations by modelling the stress tensor (and its time evolution). 

Theoretical advances in recent years have focused on constructing an effective field theory of large-scale structure (EFTofLSS) \cite[e.g.][]{baumann_2012, carrasco_2012, carrasco2014} that attempts to extend the reach of perturbation theory to larger wavenumbers. An effective theory is an approximate description of a physical system. It assumes the existence of a hierarchy of scales characterising the system it seeks to describe. Given such a hierarchy, the physics on short-wavelength (UV) scales may be described separately from that on long-wavelength (IR) scales. A UV-complete theory captures the physics on all scales, while the EFT describes only the IR scales, making no claims about the UV physics. The coupling between scales is accounted for by using coefficients that are measured from the UV-complete theory, or treated as free parameters. The starting assumption of the EFTofLSS is that there exists a length scale at which perturbation theory breaks down. This `non-linear scale' is usually described in terms of the associated wavenumber $k_\mathrm{NL}$ and approximately corresponds to the typical distance crossed by dark-matter particles until a given epoch. The goal of the EFTofLSS is to build a perturbative expansion for summary statistics of observable quantities (e.g. their correlation functions or power spectra) organised in powers of $k/k_\mathrm{NL}$ so that it converges in the perturbative regime $k\ll k_\mathrm{NL}$. The IR scales are described by modifying Eulerian\footnote{An effective theory in Lagrangian space can also be constructed \cite{porto2014, vlah2015lagrangian}.} standard perturbation theory (SPT, see e.g. \cite{bernardeau2002} for a review). Assuming that the UV and IR scales are well-separated, the EFTofLSS proceeds in the following way:
\begin{enumerate}
        \item {COARSE-GRAINING.} The Vlasov equation is smoothed over a length scale much larger than the non-linear scale. Following the convention in the literature, we characterise this smoothing scale in terms of its corresponding wavenumber $\Lambda \ll k_\mathrm{NL}$. As routinely done in fluid mechanics, a macroscopic model is obtained by taking the system of moment equations and truncating it at a suitable order. A cornerstone of the EFTofLSS is that the Vlasov hierarchy for the coarse-grained system (which only includes perturbations with wavenumber $k \ll \Lambda \ll k_\mathrm{NL}$) can be safely truncated at second order and is thus described by two `fluid' equations. The key idea here is that, due to the finite age of the Universe and their relatively low speeds, dark-matter particles did not have the time to cross long distances. From scaling arguments, it follows that the length scale associated with $k_\mathrm{NL}$ plays a similar role as the mean free path in a neutral gas dominated by collisions, i.e. higher-order moments of the phase-space density are suppressed by powers of $k/k_\mathrm{NL}$ \cite{baumann_2012}. However, contrary to the case of the rarefied gas (where the suppression is due to the frequent molecular collisions), the 
        higher-order moments in the EFTofLSS simply did not have time to develop within the age of the Universe. The time evolution of the coarse-grained system is thus regulated by the continuity equation and the Cauchy momentum equation containing an effective stress tensor $\boldsymbol{\tau}$. A number of arguments indicate that virialised structures decouple completely from the long-wavelength theory and do not alter the expansion history of the background Universe \cite{baumann_2012}. The largest contribution to the effective stress tensor is thus expected from scales that have recently turned non-linear but did not yet develop into virialised structures (i.e. Fourier modes with $k \simeq k_\mathrm{NL}$).
            
        \item {INTEGRATING OUT SMALL-SCALE FLUCTUATIONS.} In a specific realisation, the effective stress tensor depends on both the long-wavelength perturbations with $k\leq\Lambda$ (which alter the particle trajectories through tidal effects) and the short-wavelength perturbations with $k>\Lambda$. In order to eliminate the latter dependence, the EFTofLSS computes the evolution of the Fourier modes with $k\ll \Lambda$ by replacing $\boldsymbol{\tau}$ with its conditional average $\langle \boldsymbol{\tau} \rangle$ which characterises the `typical' influence of the small-scale modes on a patch of size $\Lambda^{-1}$ with fixed long-wavelength modes. This way, an effective theory that only includes long-wavelength modes is obtained.
                
        \item {STOCHASTIC TERMS.} In each realisation, the difference $\mathbf{J}=\boldsymbol{\tau}-\langle \boldsymbol{\tau} \rangle$ provides a source of noise in the evolution of the large-scale perturbations with respect to the averaged case. This stochastic contribution is not necessarily small with respect to $\langle \boldsymbol{\tau} \rangle$ but, at leading order,
        generates corrections that should be uncorrelated with the long-wavelength modes and therefore suppressed in averaged statistics like the power spectrum of density fluctuations for $k\ll \Lambda$. On the other hand, the auto-correlation of the stochastic terms does not vanish and it is often assumed to be Poisson-like on scales comparable with $k_\mathrm{NL}$ \cite{carrasco_2012}.   
        
        \item {PERTURBATIVE EFT.} 
        Since the smoothed fluid properties (viz. fluctuations in the matter density and velocity potential) lie, by construction, in the quasi-linear regime, it makes sense to study their evolution perturbatively. In order to solve the non-linear equations regulating the effective fluid, $\langle \boldsymbol{\tau} \rangle$ is expanded in terms of the long-wavelength fields and their spatial derivatives. All terms allowed by the symmetries of the problem (i.e. of the UV-complete theory) are considered. This set of interaction terms is ordered by importance using a  power-counting technique which yields a systematic expansion in $k/\Lambda$ or $k/k_\mathrm{NL}$. A finite number of relevant interaction terms appear at a given order. At the lowest one, the coefficients of the expansion can be interpreted as an effective background pressure, a speed of sound, and bulk- and shear-viscosity coefficients, in analogy with the phenomenological stress tensor for an imperfect Newtonian fluid (which leads to the Navier-Stokes equation). These time-dependent but spatially homogeneous coefficients are determined by the short-range physics and the coarse-graining scale $\Lambda$. They quantify the UV-IR coupling of the perturbations and are not predictable\footnote{A strong simplifying assumption which is regularly made in the literature is that their time dependence can be inferred in perturbation theory \cite[e.g.][]{carrasco_2012, hertzberg2014effective}.} within the effective theory. The EFTofLSS differs from SPT because of the presence of these additional terms in the fluid equations.
        
        \item {PERTURBATIVE SOLUTION.}
        Power counting suggests that the EFT corrections to the SPT solutions only appear at non-linear order. The fluid equations of the perturbative EFT are thus solved iteratively by expanding the fields around the linear solution (as in SPT). Diagrammatic techniques analogous to Feynman diagrams can be employed to evaluate correlation functions of the long-wavelength perturbations. Tree diagrams provide the leading-order solution while higher-order corrections are associated with loop diagrams and require integrations over the wavenumber. The coarse-graining scale $\Lambda$ acts as a cutoff of the loop integrals (smooth or sharp depending on the filter used to isolate the long-wavelength part of the fields). It turns out that some of the loop corrections in SPT show substantial variations with $\Lambda$ and are thus said to be UV-sensitive. This strong dependence of the IR theory on the UV degrees of freedoms likely reflects the failure of the SPT macroscopic equations on small scales (i.e. the loops integrate the SPT solutions for the fields in a regime in which they are not valid). The EFTofLSS addresses this issue by choosing a renormalisation scheme for the IR-UV couplings.
                
        \item {RENORMALISATION.} 
        The UV sensitivity of the SPT loops can be absorbed into the parameters of the effective theory order by order as follows. The EFT expressions for the correlation functions of the long-wavelength perturbations contain a number of coefficients\footnote{\label{footnote:non_locality}Further complexity is introduced by the fact that while UV and IR scales are well separated in space, the corresponding perturbations evolve on similarly long timescales, comparable with the Hubble time (before virialisation). This implies that the effective theory is non-local in time \cite[e.g.][]{carrasco_2012, carroll2014, baldauf2015bispectrum}. As a consequence, the coefficients $c(\Lambda, a)$ that are derived from the same effective-fluid parameter (e.g. the effective speed of sound) do not coincide at different perturbative orders as they correspond to time integrals containing different kernels.} $c(\Lambda, a)$ deriving from the expansion of $\langle \boldsymbol{\tau} \rangle$ in terms of the smoothed fields. These `Wilsonian' coefficients are re-written as $c(\Lambda, a)=c_\mathrm{ren}(a)+c_\mathrm{ctr}(\Lambda, a)$ where the counterterm parts $c_\mathrm{ctr}(\Lambda, a)$ are meant to regularise the behaviour of the loop results in SPT as a function of $\Lambda$ (i.e. cancel out their UV sensitivity), while the renormalised coefficients $c_\mathrm{ren}(a)$ are independent of $\Lambda$. In order for the cancellation to work, the loop and counterterm corrections must have the same dependence on $k$ (in the limit $k\to 0$), $\Lambda$ and time (but, in practice, the last two equalities are assumed to hold \cite[e.g.][]{carrasco_2012, hertzberg2014effective, ivanov2022effective}). After renormalising the coupling coefficients so that the EFT predictions for $k\ll k_\mathrm{NL}$ do not depend on $\Lambda$, the limit $\Lambda \to \infty$ is taken. Although one could invoke that UV physics on virialised scales is likely to be decoupled from the IR scales, the rationale behind this last operation is just that the final expressions become simpler as the contributions from higher-derivative terms get suppressed as $k/\Lambda$. This is a mathematical trick; the EFT, in any case, is valid only for $k\ll k_\mathrm{NL}$. After renormalising the coupling coefficients and taking the limit $\Lambda \to \infty$, all physical observables (evaluated at some fixed perturbative order) are unambiguously defined within some finite accuracy (i.e. up to corrections suppressed by some  power of $k/k_\mathrm{NL}$), once expressed in terms of the $c_\mathrm{ren}(a)$.
        
        \item {UV MATCHING.} 
        The unknown coefficients of the effective theory (at a given perturbative order) can be treated as free parameters that are determined by matching a statistic (e.g. the power spectrum) to measurements from observations or numerical simulations (which provide an approximation to the UV-complete theory) at some scale $k_\mathrm{match}\ll k_\mathrm{NL}$ (or by fitting the model to the data with $k<k_\mathrm{match}$). 
\end{enumerate}

Following this procedure, the EFTofLSS allows a perturbative treatment of the Fourier modes with $k \ll k_{\mathrm{NL}}$, while taking into account the effect of higher-$k$ modes. This framework has extended the predictive power of the theory to larger wavenumbers and/or later times (compared to SPT) for the $\Lambda$CDM matter power spectrum and bispectrum \cite[e.g.][]{carrasco_2012, pz_2013, carrasco20142, porto2014, senatore2015ir, baldauf2015bispectrum, angulo2015one}. For example, the EFTofLSS extends the $k$-reach of the models for the matter power spectrum by a factor of 2--3 \cite{alkhanishvili2022reach}. The formalism has also incorporated biased tracers \cite[e.g.][]{assassi2014renormalized, senatore2015bias, angulo2015statistics, mirbabayi2015biased, donath2020biased} (see also \cite{mcdonald2006clustering, mcdonald2009clustering, werner2020renormalization}).

Broadly, EFTs come in two flavours. The literature on the EFTofLSS follows a `bottom-up' construction consistent with the symmetries of the system without reference to the underlying UV-complete theory. This provides justification for introducing additional free parameters in the theoretical model for summary statistics of the large-scale structure. In contrast, a `top-down' construction starts from a UV-complete theory and integrates out degrees of freedom above the cutoff scale $\Lambda$. The EFT coefficients can be calculated directly from the UV-complete theory without free parameters. The two constructions result in the same dynamical equations, and differ only in the method of estimation of the coefficients. Thus, in a consistent theory, their predictions should agree with each other. For the EFTofLSS, $N$-body simulations provide an approximate UV-complete theory. For example, reference \cite{carrasco_2012} compares the bottom-up and top-down approaches in a three-dimensional universe with a $\Lambda$CDM cosmology, and finds them to be in agreement within large error bars. However, this was shown for a fine-tuned range of separations and only at a single fixed time. Reference \cite{mw_2016} specialises to the one-dimensional case of interacting sheets with CDM-like initial power spectra, and find similar agreement at a fixed time, while assuming a time-dependence for the coefficients.

Our main motivation in this work is to design the simplest universe in which the bottom-up and top-down constructions can be compared, and run controlled tests to assess the agreement between the two. Similar to \cite{mw_2016}, we study the non-linear growth of density perturbations with planar symmetry. However, we specialise to an Einstein-de Sitter (EdS) universe and consider much simpler, deterministic initial conditions, viz., two superimposed cosinusoidal perturbations with substantially different wavelengths. Our aim is to understand the impact of the short-wavelength perturbation on the long-wavelength one. The simplicity of the setting allows us to measure the time evolution of the EFT coefficients, in contrast to previous studies which perform their calculations at a fixed epoch.

The paper is structured as follows. In Sections~\ref{sec:setup} and \ref{sec:simulations}, we make our setup precise and introduce our $N$-body simulations, respectively. In Section~\ref{sec:topdown}, we measure the IR-UV interaction terms from the simulations. These coefficients are then used in Section~\ref{sec:results} to compare the next-to-leading order perturbative expressions for the matter power spectrum in SPT and EFTofLSS to the simulations. Finally, in Section~\ref{sec:summary}, we summarise our results.

\section{Setup}\label{sec:setup}
We build a toy model in which two linear cosinusoidal density perturbations with planar symmetry grow in a background EdS universe. In such a system, the overdensity develops singularities at shell crossing, while the gravitational acceleration remains bounded at all times \cite{colombi2015vlasov, taruya2017post, rampf2023renormalization}. Our goal is to study how the presence of an initially high-amplitude, short-wavelength fluctuation influences the evolution of a low-amplitude, long-wavelength one. We consider a box of comoving side $L$ with periodic boundary conditions. Thus, the linear overdensity is of the following form
    \begin{equation}\label{eq:in_den}
        \delta_\mathrm{lin}(x, a=1) = A_{1}\cos(k_{1}x - \varphi_{1}) + A_{2}\cos(k_{2}x - \varphi_{2})\;,
    \end{equation}
where our time variable is the cosmic expansion factor $a$ of the background, $A_i\in \mathbb{R}^+$, $k_i=n_i\,k_\mathrm{f}$ with $n_i\in \mathbb{N}^+$ and $k_\mathrm{f}=2\pi/L$, and $\varphi_{i}\in [0,2\pi)$. Due to the planar symmetry of the perturbations, $\delta_\mathrm{lin}$ only depends on the Cartesian spatial coordinate $x$. The overdensity in Eq.~(\ref{eq:in_den}) contains four Fourier modes,
    \begin{equation}
         \delta_\mathrm{lin}(x, a=1) = \frac{1}{2}
         \left[A_1\,\mathrm{e}^{-\mathrm{i}\varphi_1}\,\mathrm{e}^{\mathrm{i}k_1x}
         +A_1\,\mathrm{e}^{\mathrm{i}\varphi_1}\,\mathrm{e}^{-\mathrm{i}k_1x}
         +A_2\,\mathrm{e}^{-\mathrm{i}\varphi_2}\,\mathrm{e}^{\mathrm{i}k_2x}
         +A_2\,\mathrm{e}^{\mathrm{i}\varphi_2}\,\mathrm{e}^{-\mathrm{i}k_2x} \right]\;,
    \end{equation}
with their amplitudes paired so that $\delta_\mathrm{lin}$ is real valued. 

In order for $\delta_\mathrm{lin}$ to represent a realisation of a Gaussian random field, the phases $\varphi_i$ must be independently drawn from a uniform probability distribution within $[0,2\pi)$ and the coefficients $A_i/2$ must be independently drawn from a Rayleigh distribution with scale parameter (i.e. its mode) equal to $\sqrt{P_\mathrm{lin}(k_i,a=1)}$ with $P_\mathrm{lin}$ the power spectrum of the random field. However, we do not consider a Gaussian random field in this work. Rather, we set $A_1=0.05$ for the long-wavelength perturbation with $n_1=1$, and $A_2=0.5$ for the short-wavelength one with $n_2=11$. We further set $\varphi_{1} = \pi$, and consider eight different values for $\varphi_{2}$ equispaced between 0 and $2\pi$. The purpose of these different initial conditions is to create an ensemble for which the linear short-wavelength modes average to zero.

\section{\texorpdfstring{$N$}{N}-body simulations}\label{sec:simulations}
We follow the non-linear evolution of the perturbations in Eq.~\eqref{eq:in_den} with the publicly available\footnote{\href{https://bitbucket.org/ohahn/cosmo_sim_1d/}{https://bitbucket.org/ohahn/cosmo\_sim\_1d/}} $N$-body code \texttt{cosmo\_sim\_1d} which is based on an exact force calculation and uses a drift-kick-drift symplectic integrator \cite{rampf2021unveiling}. We consider $N_\mathrm{p}=125,000$ equal-mass particles (representing two-dimensional slabs) and use the Zel'dovich solution \cite{zel1970} to generate initial conditions\footnote{Note that the Zel'dovich solution is exact in 1D before shell crossing, removing the need to start the simulations at an earlier epoch.} at $a=0.5$ by displacing the particles from a uniform grid. At this time, the amplitudes of the linear overdensity in the short- and long-wavelength perturbations are (respectively) 0.25 and 0.025.

To calculate the spatial dependence of the moments and cumulants from the $N$-body particle data, we interpolate the particles onto a regular grid with $10,000$ grid points. Since all properties of our system scale with the box size $L$, we express all length scales in units of $L$, wavevectors in units of $k_{\mathrm{f}}$, and velocities in units of $H_{0}L$ (where $H_{0}$ is the Hubble constant of the background). We label the mass of the $N$-body particles as $M_{\mathrm{p}}\propto N_{\mathrm{p}}^{-1}$. 
\begin{figure}
     \centering
     \begin{subfigure}[b]{0.32\textwidth}
        \centering
        \includegraphics[scale=0.32]{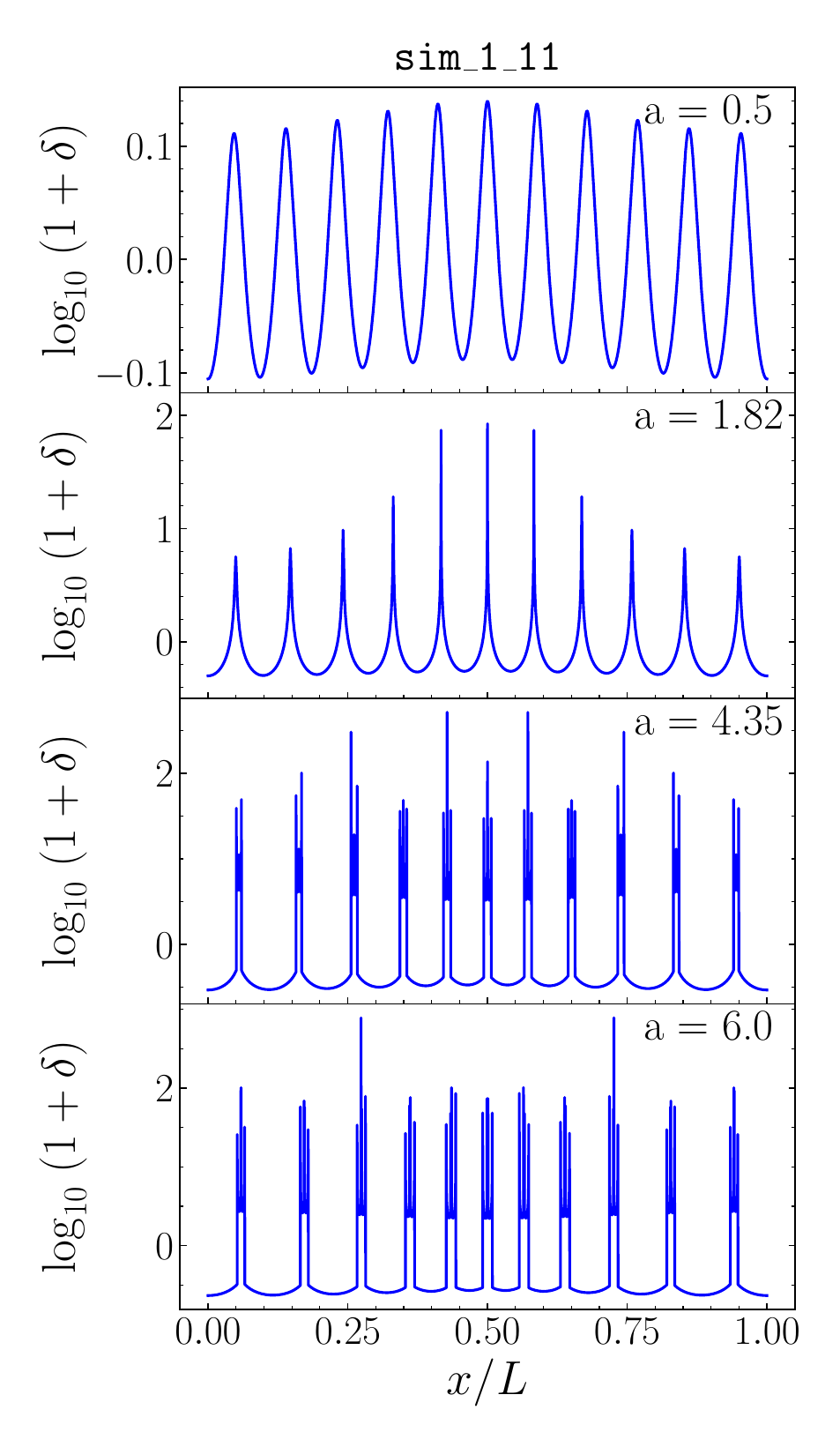}
     \end{subfigure}
     \begin{subfigure}[b]{0.32\textwidth}
        \centering
        \includegraphics[scale=0.32]{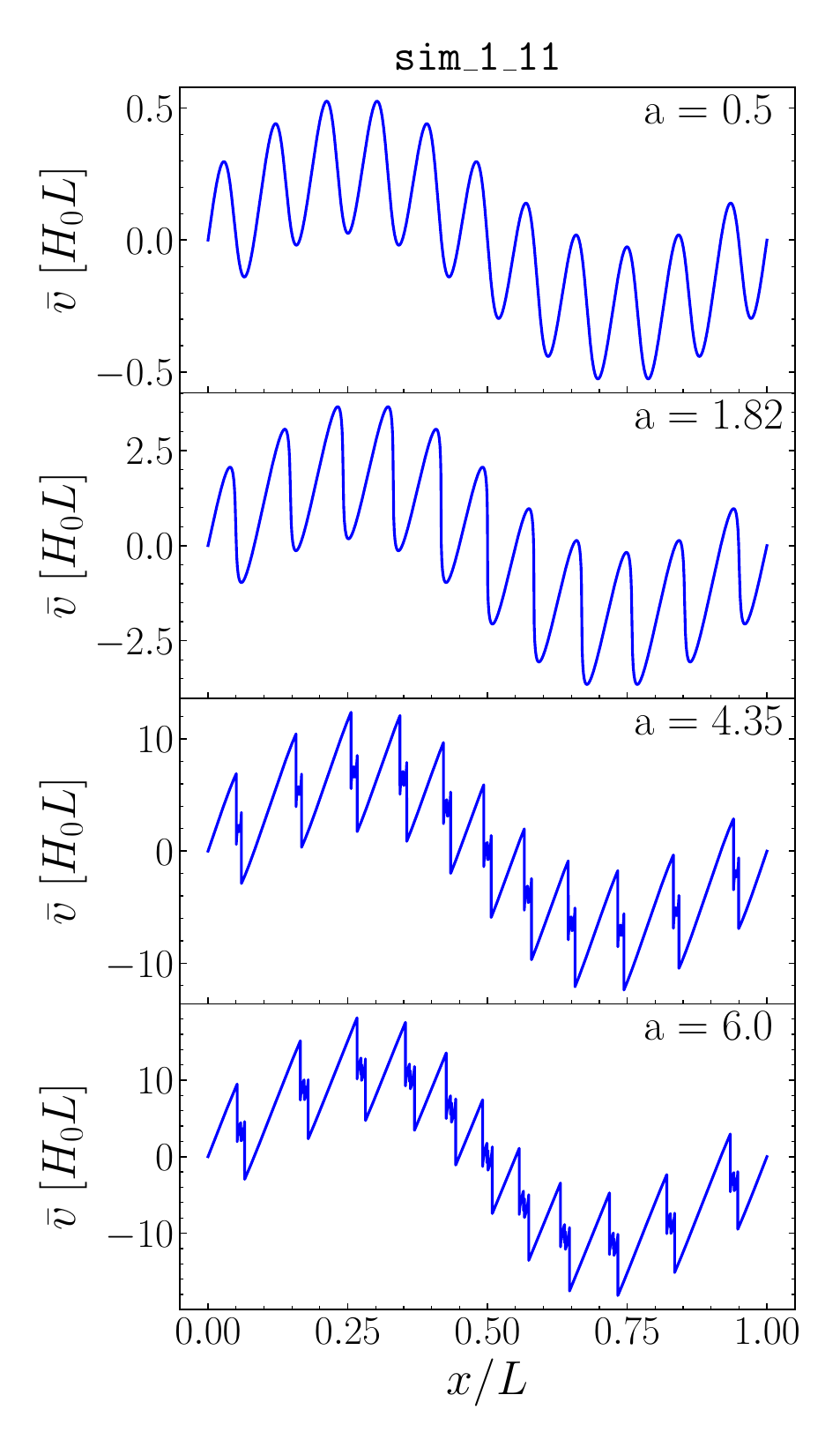}
     \end{subfigure}
     \begin{subfigure}[b]{0.32\textwidth}
        \centering
        \includegraphics[scale=0.32]{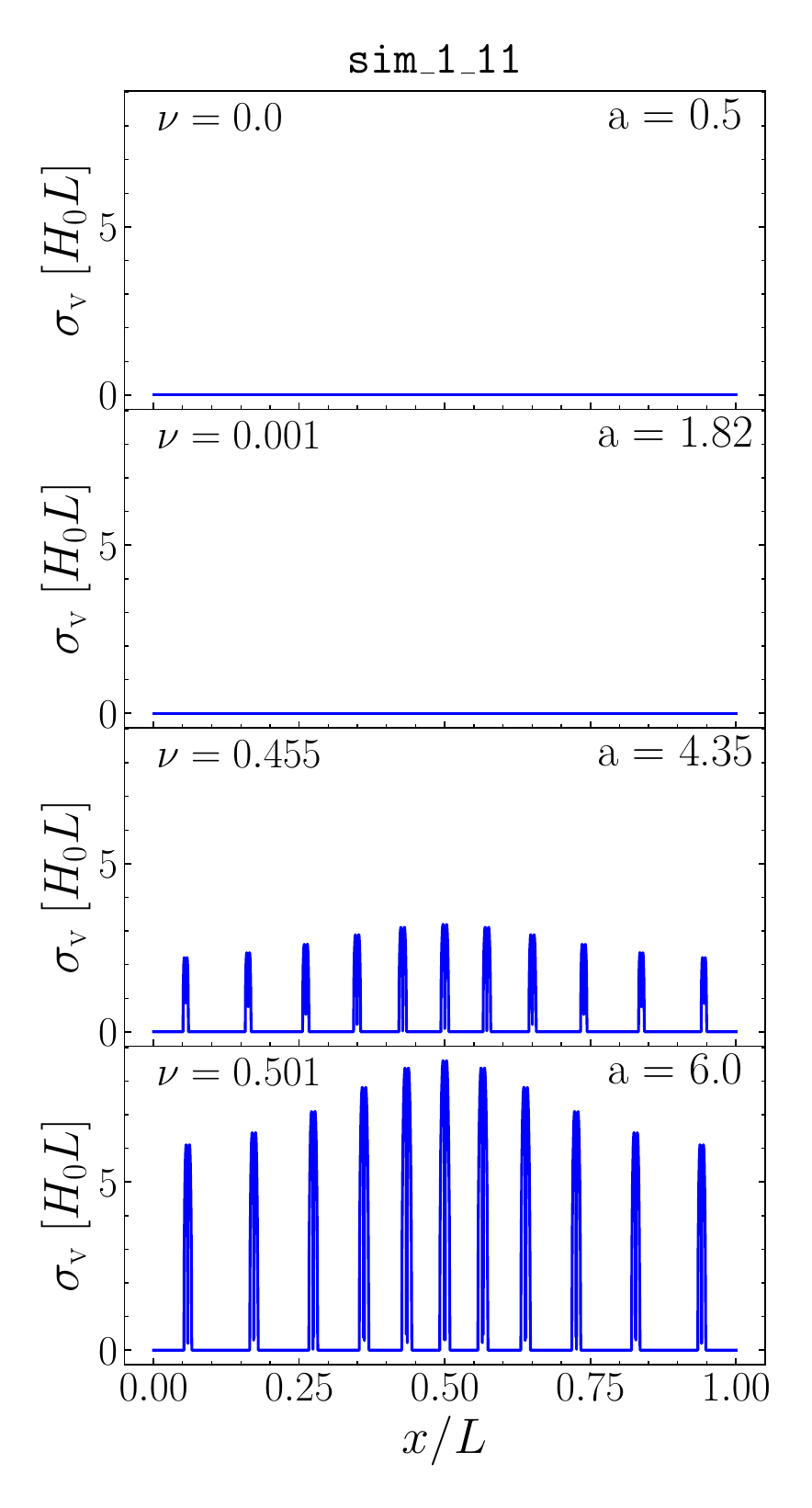}
     \end{subfigure}
    \caption{Time evolution of the overdensity, mean velocity, and velocity dispersion for \texttt{sim\_1\_11}. Shell crossing first occurs at $a=1.82$, followed by the formation of 1D halos.
    In the right panel, we indicate the fraction of particles located in multi-stream regions, $\nu$.}
    \label{fig:field_ev}
\end{figure}

\begin{figure}[t!]
        \centering
        \includegraphics[scale=0.6]{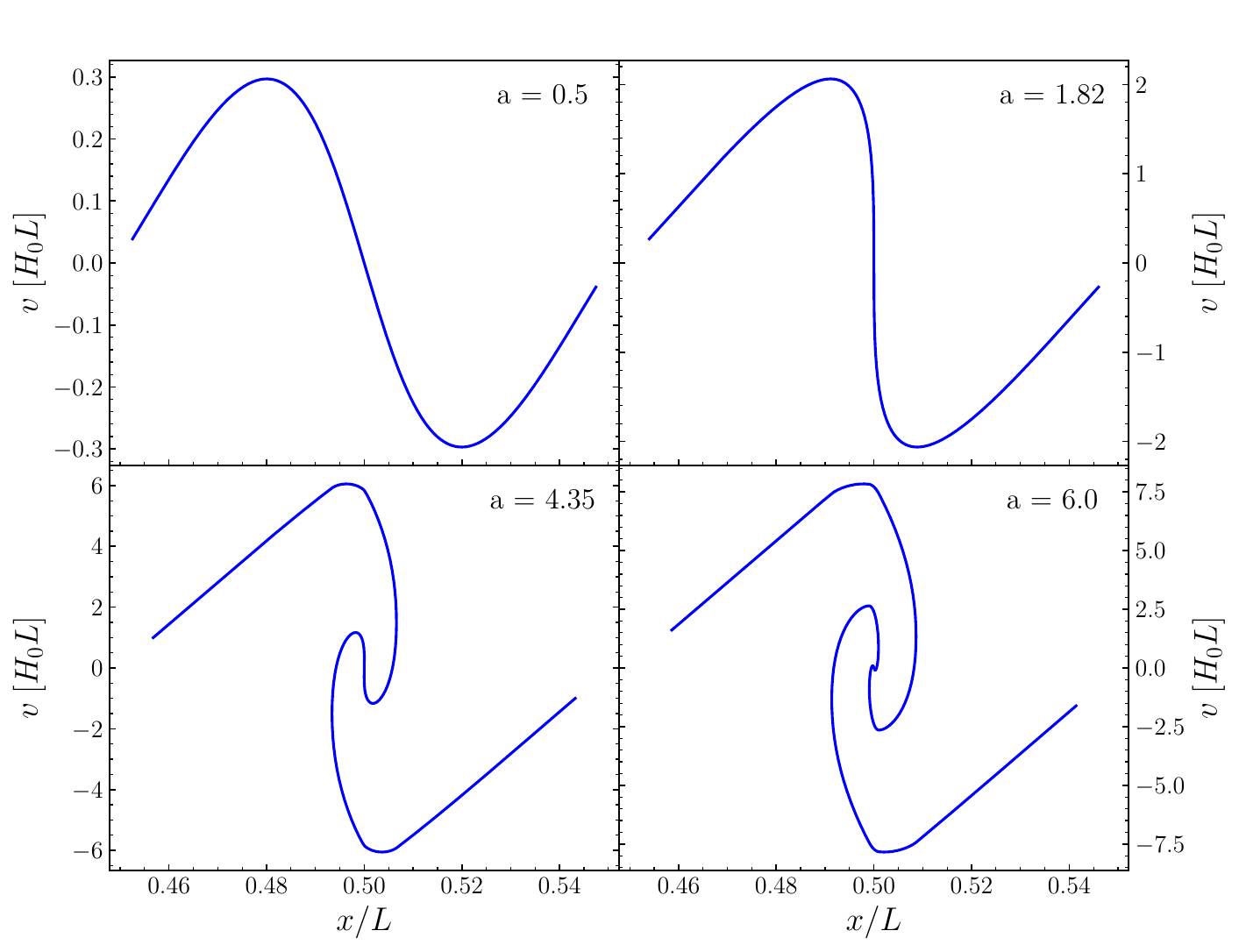}
        \caption{The particle distribution in phase-space for \texttt{sim\_1\_11}. We zoom-in on the innermost structure to show its interior more clearly. The different panels refer to
        (in order of increasing $a$) the beginning of the simulation, the first shell crossing, the second shell crossing, and the last snapshot of the simulation.}
        \label{fig:phase_space}
    \end{figure}

We name our primary simulation (with $\varphi_2=\pi$) \texttt{sim\_1\_11}, referring to the non-zero modes in the linear overdensity. The left panel of Figure~\ref{fig:field_ev} shows the evolution of the overdensity field for this simulation. At $a = 0.5$, we can see the 11 peaks that reflect the short-wavelength mode ($n_{2} = 11$) in the overdensity field. The central peak has the largest amplitude, while peaks on either side are progressively weaker. This is due to the long-wavelength mode. At later times, these initial peaks are the sites of formation of 1D halos (3D pancakes). The first shell crossing occurs at $a = 1.82$, leading to a multi-stream structure (the phase-space distribution of the innermost halo is shown in Figure~\ref{fig:phase_space}). The presence of this region signals the breakdown of the assumption of zero velocity dispersion which characterises SPT. The multi-stream structure first expands due to the momentum of the streams, but gravity slows this expansion, and eventually leads to re-collapse with a second set of shell crossings starting at $a = 4.35$. These are visible as spikes within the 1D halos in the bottom row of Figure~\ref{fig:field_ev}. We also note that the halos cluster towards the central regions of the box, because of the long-wavelength perturbation.

The same evolution can be traced in the middle panel of Figure~\ref{fig:field_ev}, where we plot the mean peculiar velocity at fixed Eulerian position. Before shell crossing occurs, the velocity field is single-valued (top row). At later times, we can infer the presence of multi-streaming regions from discontinuities in the mean velocity around the overdensity peaks (bottom two rows). In the right panel, we show the velocity dispersion for our system. Before shell crossing, this quantity is zero, but grows after shell crossing within the 1D halos. We also indicate the fraction of particles in multi-stream regions -- $\nu$; by $a=6$ (bottom row), $\nu \approx 0.5$.
    
\begin{figure}
        \centering
        \includegraphics[scale=0.54]{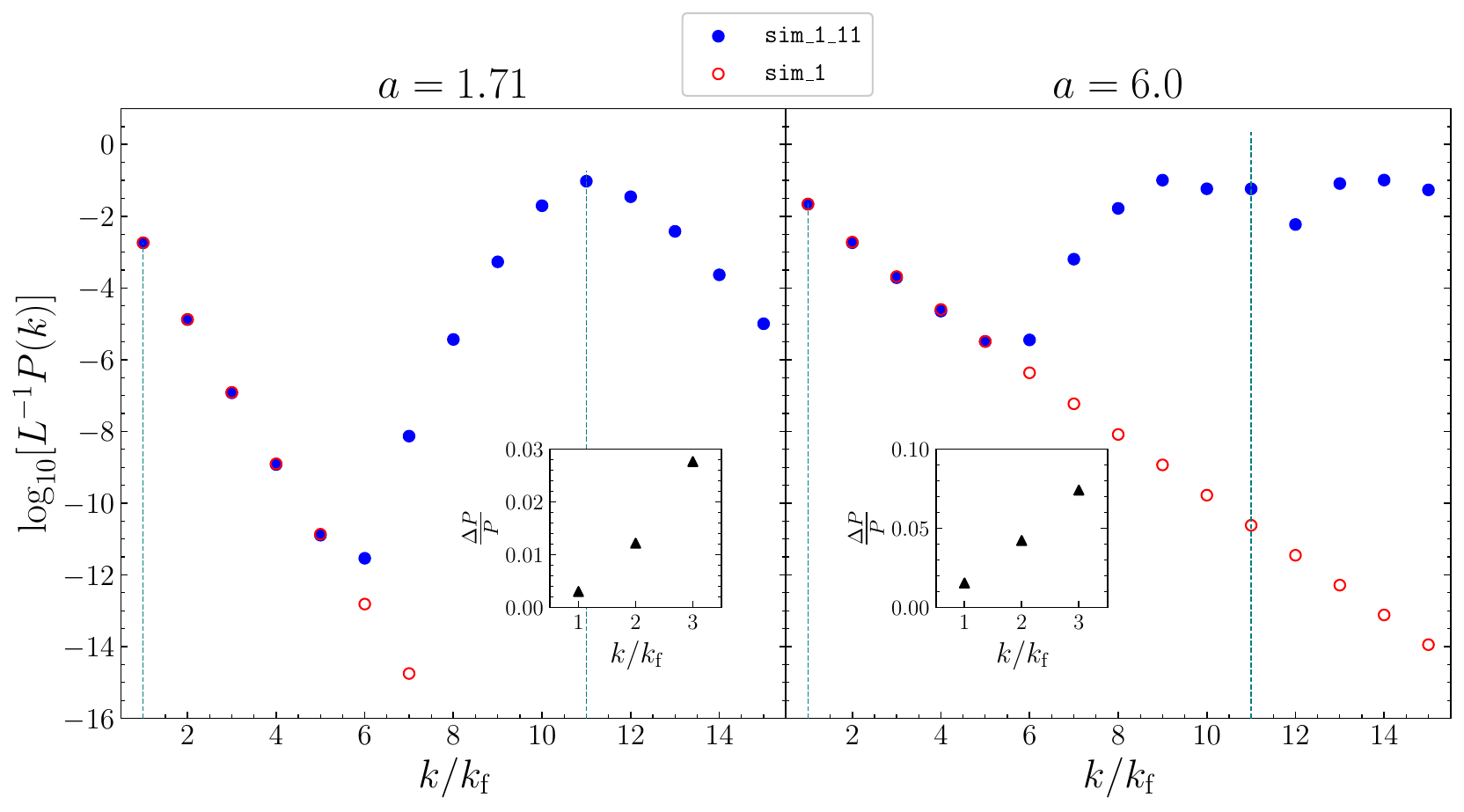}
        \caption{The $N$-body power spectra for \texttt{sim\_1} (red open circles) and \texttt{sim\_1\_11} (blue disks) at $a=1.71$ (left panel) and $a=6$ (right panel). We indicate the linear power spectra evaluated at the same expansion factors with dashed vertical lines. \emph{Inset}: the fractional difference between the spectra for $k\leq 3\,k_\mathrm{f}$, i.e. the ratio of the power measured in \texttt{sim\_1} to that measured in \texttt{sim\_1\_11} minus one.}
        \label{fig:power_spectrum}
    \end{figure}

In Figure~\ref{fig:power_spectrum}, we visualise the power spectrum of the overdensity field in \texttt{sim\_1\_11} at two epochs: $a=1.71$ (the snapshot before the first shell crossing), and $a=6$ (the final snapshot in our run). To directly quantify the impact of the high-amplitude, short-wavelength modes, we also run a simulation (hereafter \texttt{sim\_1}) with $A_2=0$ i.e., where only the long-wavelength perturbation with $n_{1}=1$ and $A_{1}=0.05$ is present in the initial conditions. For $k < 6\,k_{\mathrm{f}}$, the spectra extracted from \texttt{sim\_1} (red open circles) and \texttt{sim\_1\_11} (blue disks) both decrease nearly exponentially with $k$ as power cascades down to shorter scales from the $k=k_{\mathrm{f}}$ mode at both times. For $k \geq 6\,k_{\mathrm{f}}$, the \texttt{sim\_1\_11} spectra has excess power generated from the linear modes with $k=11\,k_{\mathrm{f}}$, while the \texttt{sim\_1} spectrum continues to decay exponentially. At $a=1.71$, the \texttt{sim\_1\_11} power peaks at $k=11\,k_{\mathrm{f}}$, while mode coupling suppresses the power at this wavenumber at $a=6$. In order to better visualise the small deviations at large scales, we plot the fractional difference between the spectra for $k \leq 3\,k_{\mathrm{f}}$ in the inset. This is the ratio between the power measured in \texttt{sim\_1} and in \texttt{sim\_1\_11}, minus one. Note that the presence of the high-frequency fluctuation in the initial conditions suppresses the power of the long-wavelength modes. At $a=1.71$, the offset between the two simulations is between 0.3 and 3\%, and, by $a=6$, the feedback from large-$k$ modes amplifies this to between 1 and 8\%. In the remainder of the paper, we will investigate whether perturbation theory and, in particular, the EFTofLSS can accurately and self-consistently model this feedback.

\section{Top-down EFT coefficients}
\label{sec:topdown}
In this section, we treat the $N$-body simulations described above as the UV-complete theory from which we derive the effective dynamics for the coarse-grained momentum moments of the dark matter particles. We start by reviewing the governing equations of the EFTofLSS.

\subsection{Transport equations for the effective fluid}\label{sec:3D_equations}
The Vlasov-Poisson system of equations provides a mean-field mathematical model for a self-gravitating collection of a large number of particles. It describes the evolution of the mean\footnote{The average can be interpreted either over an ensemble of realisations corresponding to the same macroscopic state or as a coarse-graining over (mesoscopic) phase-space patches containing a large number of particles.} one-particle phase-space density, $f(\mathbf{x},\mathbf{p},t)$, and neglects the local fluctuations due to the discreteness of the particles.

In order to study the growth of cosmological perturbations, we take the momentum moments/cumulants of the Vlasov equation to obtain transport equations for a set of `fluid-like' variables. The (physical) mass density is obtained by integrating $f$ over $\mathbf{p}$ and rescaling by the dilution factor $a^3$,
    \begin{equation}\label{eq:rho_def_3D}
        \rho(\mathbf{x},t) = \frac{m}{a^{3}(t)} \int f(\textbf{x}, \textbf{p},t)\,\dd[3]{p}=
        \rho_\mathrm{b}(t)\,[1+\delta(\mathbf{x},t)]\;,
    \end{equation}
where $m$ denotes the particle mass, $\rho_\mathrm{b}$ is the background density and $\delta$ is the density contrast. In an expanding universe, the canonical momentum conjugate to the comoving coordinate $\mathbf{x}$ is $\mathbf{p}=m\,a^{2}\,\dot{\mathbf{x}}$, where the dot indicates differentiation with respect to cosmic time $t$ and the product $\mathbf{v}=a\,\dot{\mathbf{x}}$ is commonly referred to as the peculiar velocity of the particle. Therefore, the average particle velocity is
    \begin{equation}
        \mathbf{V}(\mathbf{x},t)= \frac{\displaystyle{\int \frac{\mathbf{p}}{m\,a(t)}\,f(\textbf{x}, \textbf{p},t)\,\dd[3]{p}}}
        {\displaystyle{\int f(\textbf{x}, \textbf{p},t)\,\dd[3]{p}}}
    \end{equation}
which can be expressed as
    \begin{equation}
        \mathbf{V}(\mathbf{x},t)=\frac{\mathbf{P}(\mathbf{x},t)}{\rho(\mathbf{x},t)}
    \end{equation}
with
    \begin{equation}
        \mathbf{P}(\mathbf{x},t) = \frac{1}{a^{4}(t)} \int \mathbf{p}\,f(\textbf{x}, \textbf{p},t)\,\dd[3]{p}\;.
    \end{equation}
Similarly, the second velocity moment is given by the dyadic tensor
    \begin{equation}
        \boldsymbol{\Sigma}(\mathbf{x},t) =
        \frac{\displaystyle{\int \frac{\mathbf{p}}{m\,a(t)}\,\frac{\mathbf{p}}{m\,a(t)}\,f(\textbf{x}, \textbf{p},t)}\,\dd[3]{p}}
        {\displaystyle{\int f(\textbf{x}, \textbf{p},t)\,\dd[3]{p}}}
    \end{equation}
which can be written as
\begin{equation}
    \boldsymbol{\Sigma}(\mathbf{x},t)=
    \frac{\boldsymbol{\Xi}(\mathbf{x},t)}
    {\rho(\mathbf{x},t)}\;,
\end{equation}
with
    \begin{equation}
        \boldsymbol{\Xi}(\mathbf{x},t) =
        \frac{1}{ma^{5}(t)} \,
    \int \mathbf{p}\,\mathbf{p}\,f(\textbf{x}, \textbf{p},t)\,\dd[3]{p}\;.
    \end{equation}
This tensor is further decomposed into an unconnected streaming part and a connected dispersion part:
    \begin{equation}\label{eq:kinetic_stress_decomp_3D}
        \boldsymbol{\Xi}(\mathbf{x},t) =\rho(\mathbf{x},t)\,\mathbf{V}(\mathbf{x},t)\,\mathbf{V}(\mathbf{x},t)
        + \boldsymbol{\Gamma}(\mathbf{x},t) 
        \;.
    \end{equation}
    
The caveat in taking successive moments in this way is that the equation for the $n$\textsuperscript{th} moment depends on the $(n+1)$\textsuperscript{th} moment. Thus, a closure condition is needed to truncate the Vlasov hierarchy. SPT assumes that structure formation is driven by matter with negligible velocity dispersion (cold dark matter in the single-stream regime) and thus sets the velocity-dispersion tensor $\boldsymbol{\Gamma}$ to zero (this approximation is expected to break down with time as more and more collapsed structures form \cite{pueblas2009, buehlmann2019large}).

The governing equations are thus the mass and momentum-conservation equations coupled to the Poisson equation for the peculiar gravitational potential $\phi$:
    \begin{align}
        \dot{\delta} + \frac{1}{a}\nabla\cdot[(1 + \delta)\,\mathbf{V}] &= 0 \label{eq:spt_continuity_3D}\;, \\
        \dot{\mathbf{V}} + \frac{\dot{a}}{a}\,\mathbf{V} + \frac{1}{a}(\mathbf{V}\cdot\nabla)\mathbf{V} &=-\frac{1}{a}\nabla\phi \;,
        \label{eq:spt_euler_3D}\\
        \nabla^{2}\phi &= 4\pi G a^{2}\rho_{\mathrm{b}} \delta
        \label{eq:spt_poisson_3D} \;. 
    \end{align} 
In SPT, these equations are solved perturbatively by expanding $\delta$ and $\nabla\cdot\mathbf{V}$ (as no vorticity can be generated if $\boldsymbol{\Gamma}=0$) about the linear solutions \cite{bernardeau2002}.
    
In the EFTofLSS framework, the Vlasov-Poisson system is first smoothed with a low-pass spatial filter which removes  small-scale (but still macroscopic) fluctuations. As a second step, momentum moments of the (coarse-grained) Vlasov equation are taken. As conventional in the literature, we use the subscript $\ell$ -- short for `long-wavelength part' -- to distinguish the fluid-like variables that have been smoothed. The governing equations for the first two moments in this case are:
    \begin{align}
        \dot{\delta}_\ell + \frac{1}{a}\nabla\cdot[(1 + \delta_\ell)\,\mathbf{U}] &= 0 \label{eq:coarse_continuity_3D}\;, \\
        \dot{\mathbf{U}} + \frac{\dot{a}}{a}\,\mathbf{U} + \frac{1}{a}(\mathbf{U}\cdot\nabla)\mathbf{U} &=-\frac{1}{a}\nabla\phi_\ell - \frac{1}{a\rho_{\ell}}\nabla\cdot \boldsymbol{\tau} \;,
        \label{eq:coarse_euler_3D}\\
        \nabla^{2}\phi_\ell &= 4\pi G a^{2}\rho_{\mathrm{b}} \delta_\ell
        \label{eq:coarse_poisson_3D} \;, 
    \end{align}
where $\mathbf{U}=\mathbf{P}_\ell/\rho_\ell$ and $\boldsymbol{\tau}$ denotes the effective stress tensor which quantifies the dynamical coupling to the degrees of freedom suppressed by the smoothing procedure. The EFTofLSS closes the system of equations by modelling $\boldsymbol{\tau}$. 

Considering perturbations with planar symmetry greatly simplifies the mathematical formulation as we can pick a reference system in which all fluid functions depend only on one Cartesian coordinate, say $x^1$ (simply $x$ hereafter). Moreover, $U^i=U\,\delta^i_1$, $\tau^{ij}=\tau \,\delta^{i}_1\,\delta^{j}_1$ and $\Xi^{ij}_\ell=\Xi_\ell \,\delta^{i}_1\,\delta^{j}_1$
where $\delta^i_j$ denotes the Kronecker symbol. From now on, we refer to $\tau$ as the \emph{effective stress}.

Reference \cite{buchert2005adhesive} (see also e.g. \cite{carrasco_2012}) has shown that the effective stress naturally decomposes into two pieces
    \begin{equation}\label{eq:tau_def_1}
        \tau= \tau_\mathrm{k}+ \tau_\mathrm{g}\;,
    \end{equation}
where the kinetic part $\tau_\mathrm{k}$ accounts for the velocity dispersion within a smoothing patch
    \begin{equation}\label{eq:tau_def_2}
        \tau_\mathrm{k}= \Xi_{\ell} - \rho_{\ell}\,U^2 \;,
    \end{equation}
and the gravitational part $\tau_\mathrm{g}$ quantifies the departures from the mean-field peculiar acceleration $-\partial_x \phi_l/a$
    \begin{equation}\label{eq:tau_def_3}
        \tau_\mathrm{g} = \frac{1}{8\pi G a^{2}} \left\{\left[(\partial_{x}\phi)^{2}\right]_{\ell} - \left(\partial_{x}\phi_{\ell}\right)^{2}\right\} \;.
    \end{equation}

\subsection{Coarse-graining and non-linear scales}\label{sec:coarse-graining}
We now return to our $N$-body simulations and measure the effective stress in their snapshots at various output times using Eqs.~(\ref{eq:tau_def_1}--\ref{eq:tau_def_3}). First, we smooth all the relevant quantities with a low-pass filter defined by a kernel $W_{\Lambda}(x)$. Specifically, for a generic field $g$, we compute the convolution
    \begin{equation}
        g_{\ell}(x) = \int W_{\Lambda}(x-x') \,g(x')\,\dd{x'}\;.
    \end{equation}    
We use the FFT algorithm to perform convolutions in Fourier space, $\widetilde{g}_l(k)=\widetilde{W}_\Lambda(k)\,\widetilde{g}(k)$. To check whether our results depend on the choice of the kernel, we use both a sharp $k$-space cutoff,  $\widetilde{W}_\Lambda(k)=\Theta(\Lambda-k)$ where $\Theta$ is the Heaviside step function, and a smooth Gaussian cutoff, $\widetilde{W}_\Lambda(k)=\exp[-k^{2}/(2\Lambda^2)]$. The key difference between these two choices is that the Gaussian filter receives contributions from modes with $k>\Lambda$.

\begin{figure}[t!]
        \centering
        \includegraphics[scale=0.65]{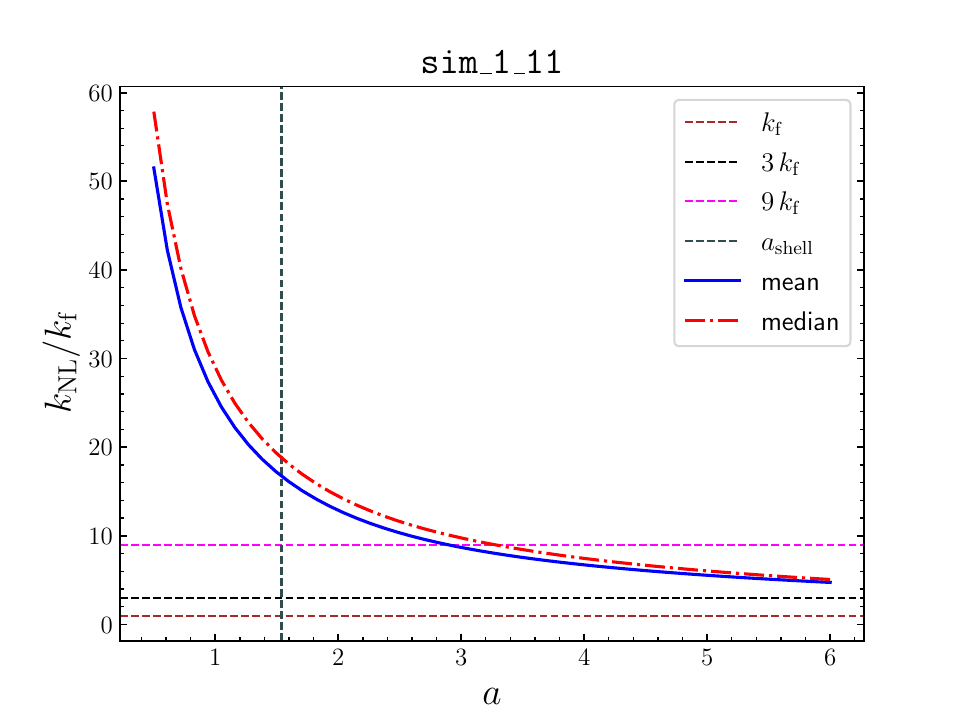}
        \caption{Time evolution of $k_{\mathrm{NL}}$ for \texttt{sim\_1\_11}. Here, $k_{\mathrm{NL}}$ denotes the wavenumber corresponding to the mean (solid blue line) or median (dash-dotted red line) particle displacement. We indicate three important scales with horizontal dashed lines: $k=k_{\mathrm{f}}$ is the scale of interest in our work, $k=3\,k_{\mathrm{f}}$ is the coarse-graining scale we mostly use, and $k=9\,k_{\mathrm{f}}$ is the smallest $k_{\mathrm{NL}}$ which satisfies the ordering of scales $k \ll \Lambda \ll k_{\mathrm{NL}}$. The vertical dashed line highlights the epoch of first shell crossing.}
        \label{fig:k_NL}
    \end{figure}

As previously mentioned, the EFTofLSS should be applicable only when the hierarchy of scales $k\ll \Lambda \ll k_\mathrm{NL}$ holds true. It is thus necessary to first get an estimate of $k_\mathrm{NL}$ in order to pick an appropriate value for $\Lambda$. We do this by taking the inverse of the mean (or median) displacement of the simulation particles (relative to their initial position at $a=0$) at each epoch. As an example, we plot the evolution of $k_{\mathrm{NL}}$ with $a$ for \texttt{sim\_1\_11} in Figure~\ref{fig:k_NL}. The non-linear wavenumber rapidly decreases with time and, at the epoch of first shell crossing (highlighted with a vertical dashed line), $k_{\mathrm{NL}}\simeq 15 \,k_\mathrm{f}$. This is a reasonable estimate given that the shell crossing is generated by the collapse of the linear perturbation with $k=11\,k_\mathrm{f}$. 

In what follows, we will be mainly concerned with the non-linear evolution of the longest perturbation in the simulation box, i.e. $k = k_{\mathrm{f}}$. Since $k_\mathrm{f}\ll k_\mathrm{NL}$ for $a<6$, it should be possible to treat the Fourier modes of $\delta$ with wavenumber $|k|=k_\mathrm{f}$ perturbatively. With the above discussion in mind, we need to pick a value of $\Lambda$ that is simultaneously much smaller than the wavenumber of the large-$k$ mode ($11\,k_{\mathrm{f}}$), and larger than the scale of interest. The latter is the largest scale in our system ($k = k_{\mathrm{f}}$). Thus, $\Lambda = 3\,k_{\mathrm{f}}$ or $4\,k_{\mathrm{f}}$ are natural choices. For most of our analysis, we will make the choice $\Lambda = 3\,k_{\mathrm{f}}$ (i.e. the sharp-$k$-space filter retains the modes with $|k|=k_{\mathrm{f}}$, $2\,k_{\mathrm{f}}$, and $3\,k_{\mathrm{f}}$) which is three times larger than the scale of interest. Thus, $\Lambda/k = 3$, and $k_{\mathrm{NL}}/\Lambda$ should be similar to ensure that the hierarchy $k\ll \Lambda \ll k_\mathrm{NL}$ holds. The EFTofLSS should then be applicable until $k_{\mathrm{NL}} = 9\,k_{\mathrm{f}}$ which corresponds to $a\approx 3$ (see Figure~\ref{fig:k_NL}). In principle, a larger value of $\Lambda$ may also be chosen, as long as it remains much smaller than $11\,k_{\mathrm{f}}$. We have checked that our results do not depend on the smoothing scale for $\Lambda \leq 8\,k_{\mathrm{f}}$.

\subsection{Effective stress}\label{sec:tau}
\begin{figure}
        \begin{subfigure}{\linewidth}
            \centering
            \includegraphics[width=\textwidth]{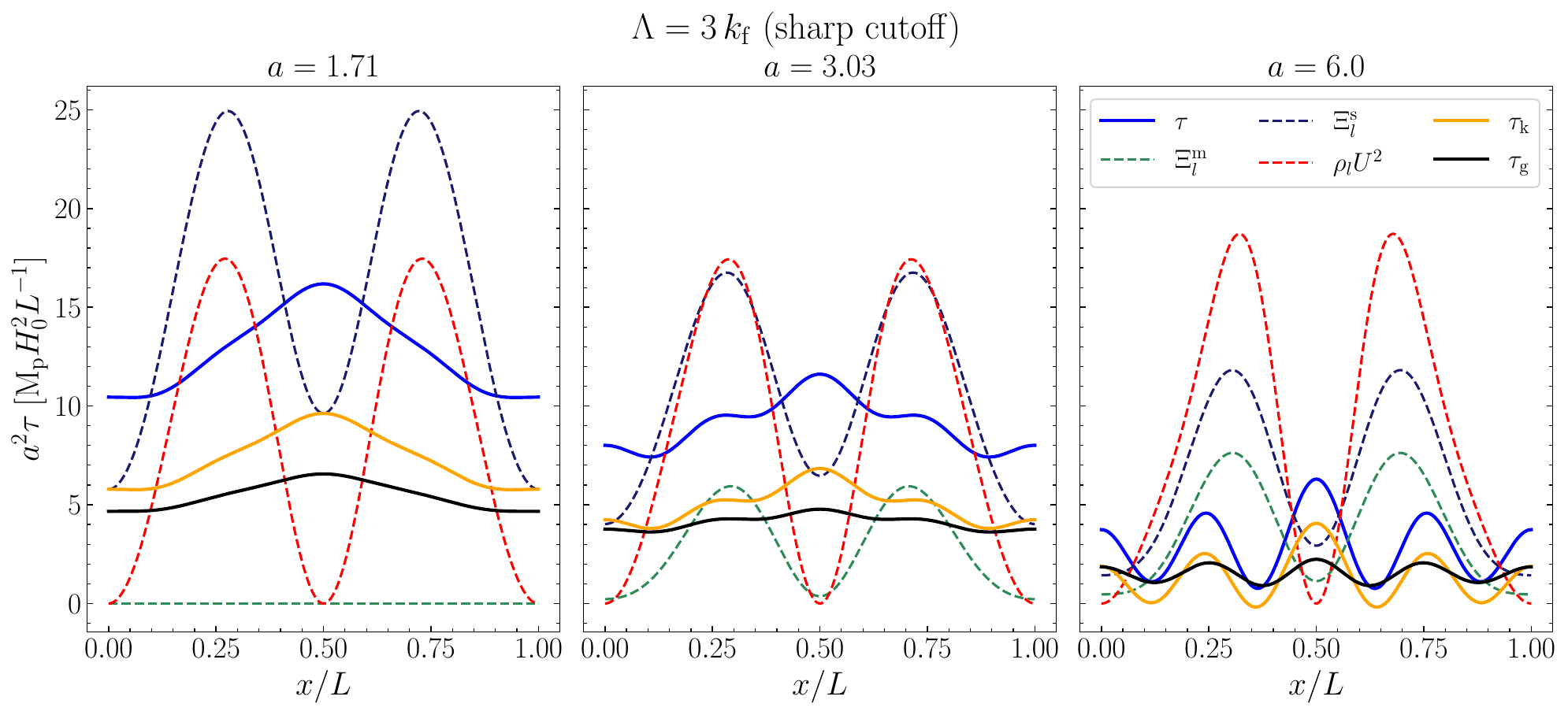}
        \end{subfigure}
        \begin{subfigure}{\linewidth}
            \vspace{0.25cm}
            \centering
            \includegraphics[width=\textwidth]{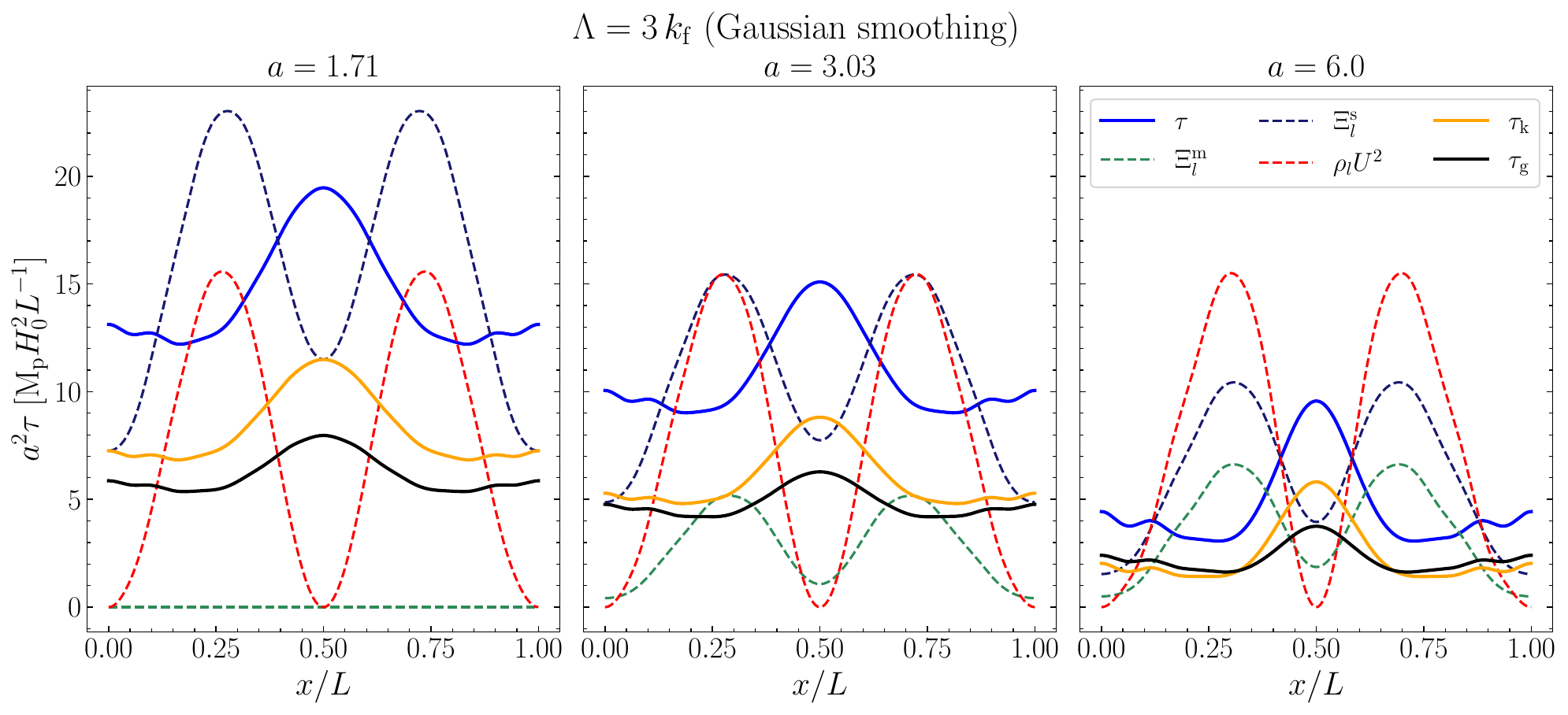}
        \end{subfigure}
        \caption{The effective stress in \texttt{sim\_1\_11} decomposed into its constituent pieces as described in Eqs. (\ref{eq:tau_def_1}--\ref{eq:tau_def_3}) for sharp cutoff (top row) and Gaussian smoothing (bottom row). The superscript over $\Xi_{\ell}$ indicates whether the contribution is made by particles within single-stream (s) or multi-stream (m) regions. }
        \label{fig:tau_decomp}
    \end{figure}

The effective stress measured in \texttt{sim\_1\_11} is shown in Figure~\ref{fig:tau_decomp} (solid blue line) together with the different contributions defined in Eqs.~(\ref{eq:tau_def_1}--\ref{eq:tau_def_3}). The coarse-grained second moment $\Xi_{\ell}$ is split into contributions from matter in the single-stream regime ($\Xi_{\ell}^\mathrm{s}$) and in the multi-streaming regime ($\Xi_{\ell}^\mathrm{m}$). At early times, when the flow is in the single-stream regime (i.e. $\Gamma=0$), we can use linear perturbation theory to predict the time dependence of the different contributions to $\tau$. Since the peculiar gravitational potential $\phi$ does not evolve with time and the peculiar velocity of the fluid scales as $a^{1/2}$, it follows that both $\tau_\mathrm{k}$ and $\tau_\mathrm{g}$ should decrease proportionally to $a^{-2}$. In Figure~\ref{fig:tau_decomp}, we therefore multiply all the components by $a^2$ to cancel out this linear evolution.

The first row displays the sharp-cutoff case. At $a=1.71$ (left panel), the flow is still in the single-stream regime, and the effective stress receives a slightly larger contribution from the kinetic part than the gravitational part. At later times, we expect the scalings discussed above to not hold, as the system becomes more and more non-linear. After shell crossing, $a^{2}\tau$ decays with time. In the middle panel ($a=3.03$), we note a non-zero contribution from the multi-stream regime via $\Xi^{\mathrm{m}}_{\ell}$. In the right panel ($a=6$), we observe oscillations in the effective stress, with four prominent peaks and valleys.

In the second row, we plot the different contributions when using a Gaussian smoothing. At all times, the total stress $\tau$ is larger in comparison to the sharp case. This excess comes from modes with $k > \Lambda$ which provide damped but non-zero contributions even at early times. Despite noticeable differences between the effective stress measured from the two smoothing schemes, the results are in rough qualitative agreement.

\subsection{Effective speed of sound}\label{sec:ctot2}
In order to close the system of equations (\ref{eq:coarse_continuity_3D}--\ref{eq:coarse_poisson_3D}), it is necessary to model the effective stress in terms of the fluid variables $\delta_\ell$ and $U$. This seems to be an impossible task as $\tau$ contains a dependence on short-wavelength modes. To address this issue, the EFTofLSS assumes that the effective stress can be decomposed into a deterministic part (which depends on $\delta_\ell$ and $U$) and a stochastic part (uncorrelated with the long-wavelength fields). The former quantifies the tidal effects of the long modes onto the particle trajectories and can be thought of as the conditional expectation value of $\tau$ over the short-wavelength perturbations at fixed long-wavelength background \cite{baumann_2012,carrasco_2012, carroll2014}, $\langle \tau\,|\,\mathrm{fixed\ background}\rangle$ (in the remainder of this paper, we denote this quantity with the symbol $\langle\tau\rangle$). However, it is not clear if and how the ensemble can be constructed in practice (e.g. using numerical simulations). In fact, this would require being able to build identical replicas of the non-linearly evolved long modes with different sets of short modes which is a formidable problem as the short modes influence the evolution of the long ones. We discuss two possible ways of addressing this issue in Sections \ref{sec:est_from_corr} and \ref{sec:fit_to_sims}.

For the moment, we assume that $\langle \tau\rangle$ exists. The next step is to expand $\langle\tau\rangle$ in terms of the long-wavelength fields. Since the characteristic time scales with which
long- and short-wavelength perturbations evolve are similar, $\langle \tau \rangle$ at a given time should depend on the entire past history of, say, $\delta_\ell$  averaged over a kernel with a width of the order of the Hubble time. This concept is expressed concisely by saying that the effective theory is `non-local in time' \cite[e.g.][]{carrasco_2012, carroll2014, baldauf2015bispectrum}. It turns out, however, that perturbative solutions can be obtained by expanding $\langle \tau\rangle$ as a sum of operators all evaluated at the same time. The non-locality of the theory will be reflected in the fact that the effective coefficients of a given operator will not coincide at different perturbative orders (see also footnote \ref{footnote:non_locality}).
This is not a concern for this paper which only considers the leading-order corrections to the SPT power spectrum. Then, we expand $\langle \tau\rangle$ in derivatives and powers of the long-wavelength perturbations $\delta_{\ell}$ and $\theta_{\ell}$ (with $\theta_{\ell} = \partial_{x}U / aH$) including all terms allowed by symmetries. Writing explicitly only the terms to linear order in  $\delta_{\ell}$ and $\theta_{\ell}$, we obtain
    \begin{align}\label{eq:tau_expansion}
        \tau = \langle \tau\rangle + J  \approx p_{b} + \rho_{\mathrm{b}} c^{2}_{\mathrm{s}} \delta_{\ell} + \rho_{\mathrm{b}}c^{2}_{\mathrm{v}} \theta_{\ell} + \ldots + J \;,
    \end{align} 
where $p_{\mathrm{b}}$ is the effective background pressure\footnote{The literature on the EFTofLSS defines $\tau$ with the opposite sign with respect to the standard convention used in fluid dynamics. Therefore, $\tau$ coincides with the pressure tensor defined in the study of non-equilibrium continua.} (which, being constant in space, has no dynamical effects), $c^{2}_{\mathrm{s}}$ is the speed of sound in the effective fluid, $c^{2}_{\mathrm{v}}$ is a (bulk-)viscosity coefficient expressed in units of speed (in doing so, we divide the viscosity by $\rho_\mathrm{b}$), the ellipses represent higher-order terms, and $J$ is the stochastic term. Note that $c^{2}_{\mathrm{s}}$ and $c^{2}_{\mathrm{v}}$ are assumed to not vary with position but are time dependent. The stochastic term is a stationary random field which varies in space and time.
    \begin{figure}
        \begin{subfigure}{\textwidth}
            \centering
            \includegraphics[width=\textwidth]{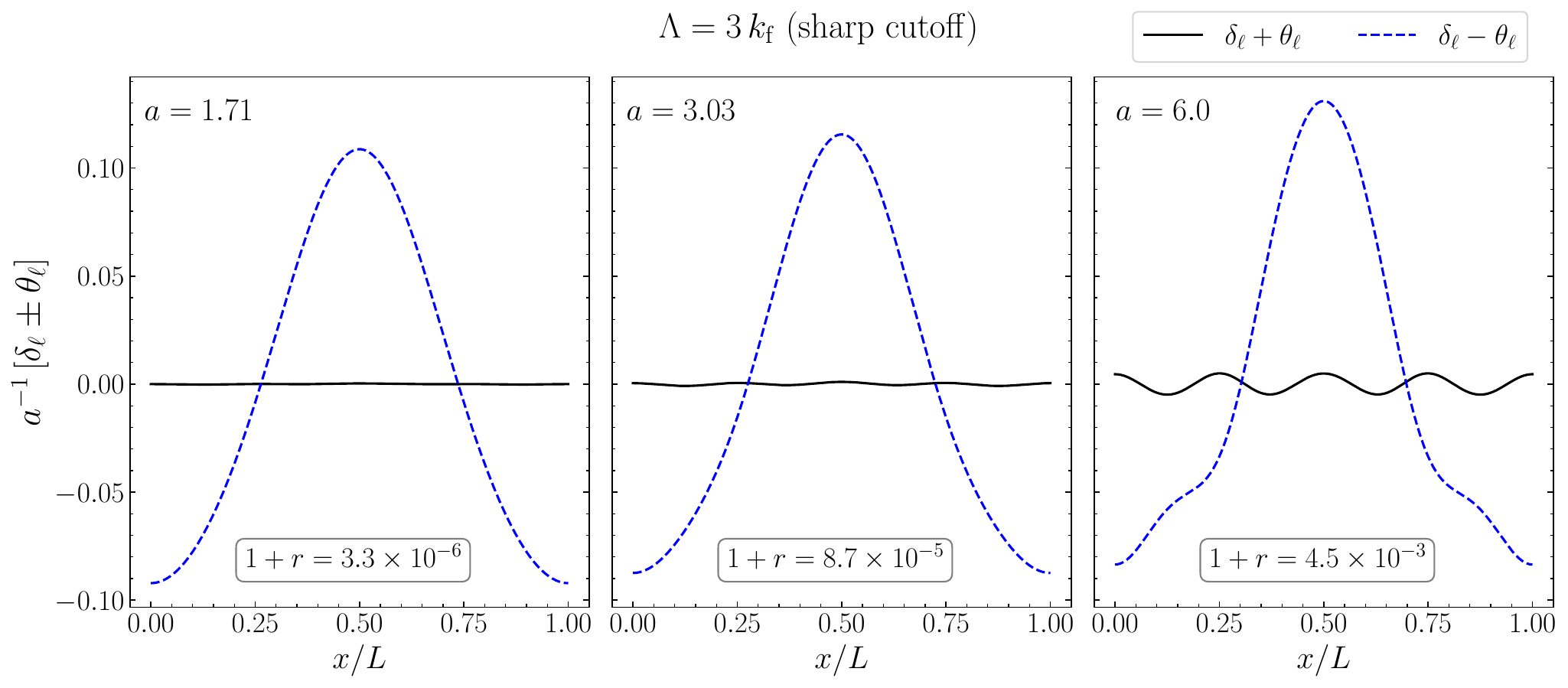}
        \end{subfigure}
        \begin{subfigure}{\textwidth}
            \centering
            \vspace{0.45cm}
            \includegraphics[width=\textwidth]{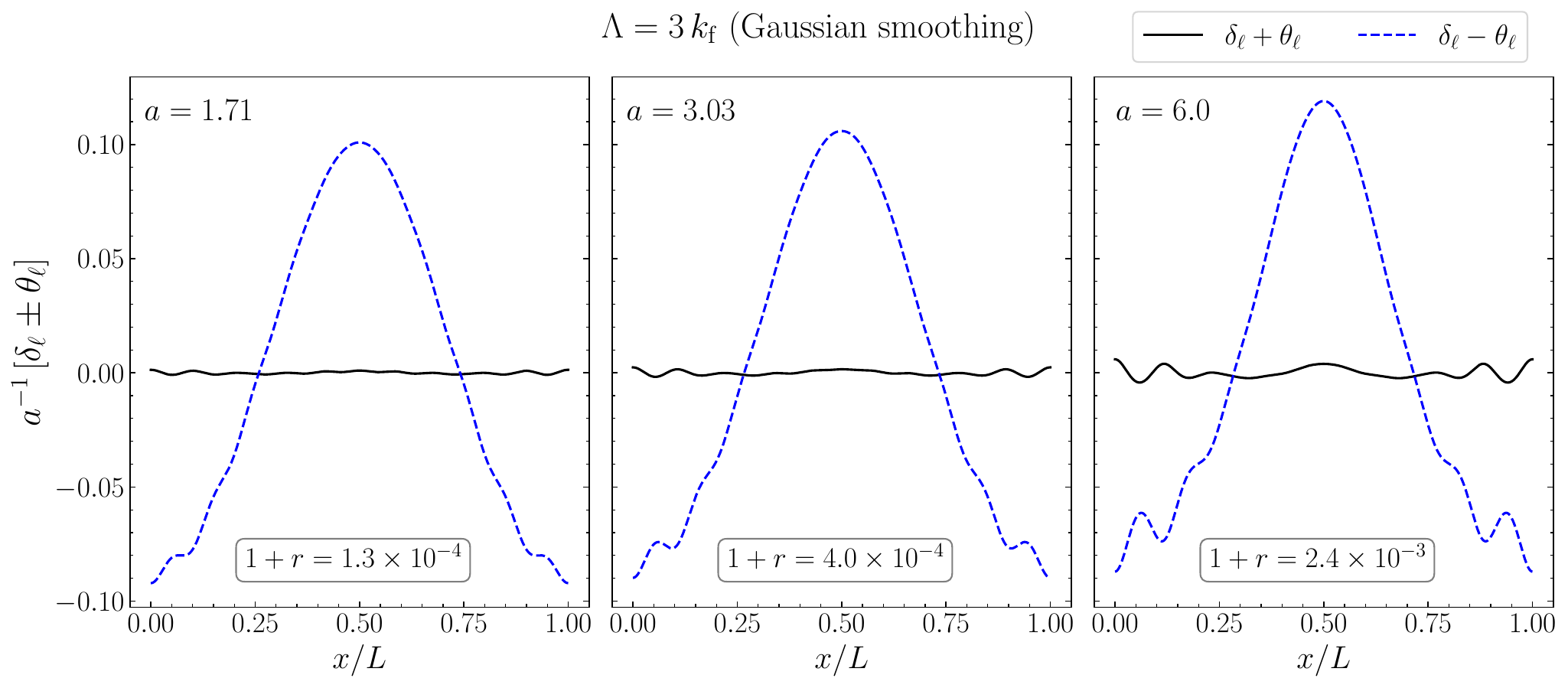}
        \end{subfigure}
        \caption{Time evolution of the fields $(\delta_\ell+\theta_\ell)/a$ (solid black) and $(\delta_\ell-\theta_\ell)/a$ (dashed blue) in \texttt{sim\_1\_11} for sharp-$k$ (top row) and Gaussian filtering (bottom row) together with $1+r$, where $r$ is the linear cross-correlation coefficient between $\delta_{\ell}$ and $\theta_{\ell}$.}
        \label{fig:delta_theta}
    \end{figure}
    
In the following subsections, we discuss different methods to estimate $c^{2}_{\mathrm{s}}$ and $c^{2}_{\mathrm{v}}$ from the simulations.

\subsubsection{Estimates from spatial cross-correlations}\label{sec:est_from_corr}
The simplest method to estimate the EFT coefficients is to assume that the leading-order expansion in Eq.~(\ref{eq:tau_expansion}) is exact and measure the spatial cross-correlation between $\tau$
and either $\delta_\ell$ or $\theta_\ell$ in a single simulation. This approach circumvents the problem of defining an ensemble in order to measure $\langle \tau\rangle$. Since the stochastic term should be uncorrelated with the long-wavelength fields, we obtain
    \begin{align}
        \langle \delta_\ell^2\rangle_x \,c^2_\mathrm{s} + \langle
        \theta_\ell\, \delta_\ell \rangle_x \, c^2_\mathrm{v}&=\langle \tau\,\delta_\ell\rangle_x/\rho_\mathrm{b}
        \label{eq:systemcs1}\\
        \langle \theta_\ell\,\delta_\ell\rangle_x\,c^2_\mathrm{s} + 
        \langle \theta_\ell^2\rangle_x\,c^2_\mathrm{v}&=\langle \tau\,\theta_\ell\rangle_x/\rho_\mathrm{b}\,
        \label{eq:systemcs2}
    \end{align}
where the symbol $\langle \dots\rangle_\mathrm{x}$ denotes averaging over different locations in space (related approaches have been used in the spatial domain by \cite{carrasco_2012}, and in the wavenumber domain by \cite{baumann_2012, mw_2016}).

At early times, when the long-wavelength perturbations are quasi-linear, $\theta_\ell\simeq -\delta_\ell$ and the system given in Eqs.~(\ref{eq:systemcs1}) and (\ref{eq:systemcs2}) is nearly degenerate. In this regime, the expansion of the effective stress reduces to 
    \begin{equation}
        \tau \approx p_{b} + \rho_{\mathrm{b}} (c^{2}_{\mathrm{s}} - c^{2}_{\mathrm{v}})\, \delta_{\ell} + \ldots+J\;,
        \label{eq:ctotexp}
    \end{equation}
as, in practice, there is only one dynamical variable. Although $\delta_\ell$ and $-\theta_\ell$ differ more and more as the long-wavelength perturbations evolve into the non-linear regime, it turns out that these fields nearly coincide at all times probed by our simulations. This is demonstrated in Figure~\ref{fig:delta_theta} where we show that $(\delta_\ell-\theta_\ell)/a$ is always substantially larger than $(\delta_\ell+\theta_\ell)/a$ when $\Lambda=3\, k_\mathrm{f}$. The linear correlation coefficient $r$ between $\delta_\ell$ and $\theta_\ell$ (also reported in the figure) is always very close to $-1$ (i.e., $1+r$ is close to zero). Fortunately, the coefficient matrix of the system (\ref{eq:systemcs1}) and (\ref{eq:systemcs2}) is numerically invertible for $a\geq 0.5$, and we can still obtain an estimate for $c_{\mathrm{s}}^{2}$ and $c_{\mathrm{v}}^{2}$. We label this estimator SC (short for spatial correlations). To cross-check that our results are not influenced by the numerical matrix inversion, we also use the following estimator for the combination $c_{\mathrm{tot}}^{2} = c_{\mathrm{s}}^{2} - c_{\mathrm{v}}^{2}$,
    \begin{equation}
        c^2_\mathrm{tot}=\frac{1}{\rho_\mathrm{b}}\frac{
        \langle \tau\,\delta_\ell \rangle_\mathrm{x}}{\langle \delta_\ell^2\rangle_\mathrm{x}}\;,
        \label{eq:cscross}
    \end{equation}
obtained from Eq.~(\ref{eq:ctotexp}). We dub this estimator SC$\delta$.

Reference \cite{mw_2016} used a similar estimator in the wavenumber domain and further assumed that the linear density perturbation $\delta_\mathrm{lin}$ provides an accurate model on large scales,
    \begin{equation}\label{eq:mw_est}
        c^{2}_{\mathrm{tot}} = 
        \frac{1}{\rho_{\mathrm{b}}}\,
        \frac{\displaystyle{\sum_{k<\Lambda} \widetilde{\tau}(k) \,\widetilde{\delta}_{\mathrm{lin}, l}(-k)}}{\displaystyle{\sum_{k<\Lambda}\left|\widetilde{\delta}_{\mathrm{lin}, l}(k)\right|^{2}}} \;,
    \end{equation}
where the tilde denotes a Fourier transform. To connect with the previous literature, we also include this estimator (which we dub the M\&W estimator) in our analysis.

\subsubsection{Estimates from least-squares fitting} \label{sec:fit_to_sims}
We can also estimate $c^2_\mathrm{s}$ and $c^2_\mathrm{v}$ by fitting the expansion in Eq.~(\ref{eq:tau_expansion}) to the simulation data. However, if $\tau$, $\delta_\ell$ and $\theta_\ell$ are measured from a single simulation, the least-squares method gives back Eqs.~(\ref{eq:systemcs1}) and (\ref{eq:systemcs2}) for the best-fitting values.

Alternatively, we can first compute $\langle\tau\rangle$ and then perform the fit. As discussed above, constructing the average $\langle \tau \rangle$ is a considerable challenge as it requires conditioning on non-linearly evolved fields. It is easier, instead, to build an ensemble of realisations in which the \emph{linear} long modes are identical and the short modes vary. This is a well-defined ensemble which can be expressed in terms of a probability distribution for the linear modes. Therefore, we compute $\langle \tau \rangle$ by averaging at fixed $x$ over the eight simulations with different values of the angle $\varphi_2$ (i.e. different phases between the $k=k_\mathrm{f}$ and $11\,k_\mathrm{f}$ modes) described in Section~\ref{sec:setup}. We note that this ensemble average is nearly identical to the $\tau$ measured from \texttt{sim\_1\_11}, especially in the sharp-cutoff case.

\begin{figure}
        \begin{subfigure}{\textwidth}
            \centering
            \includegraphics[width=\textwidth]{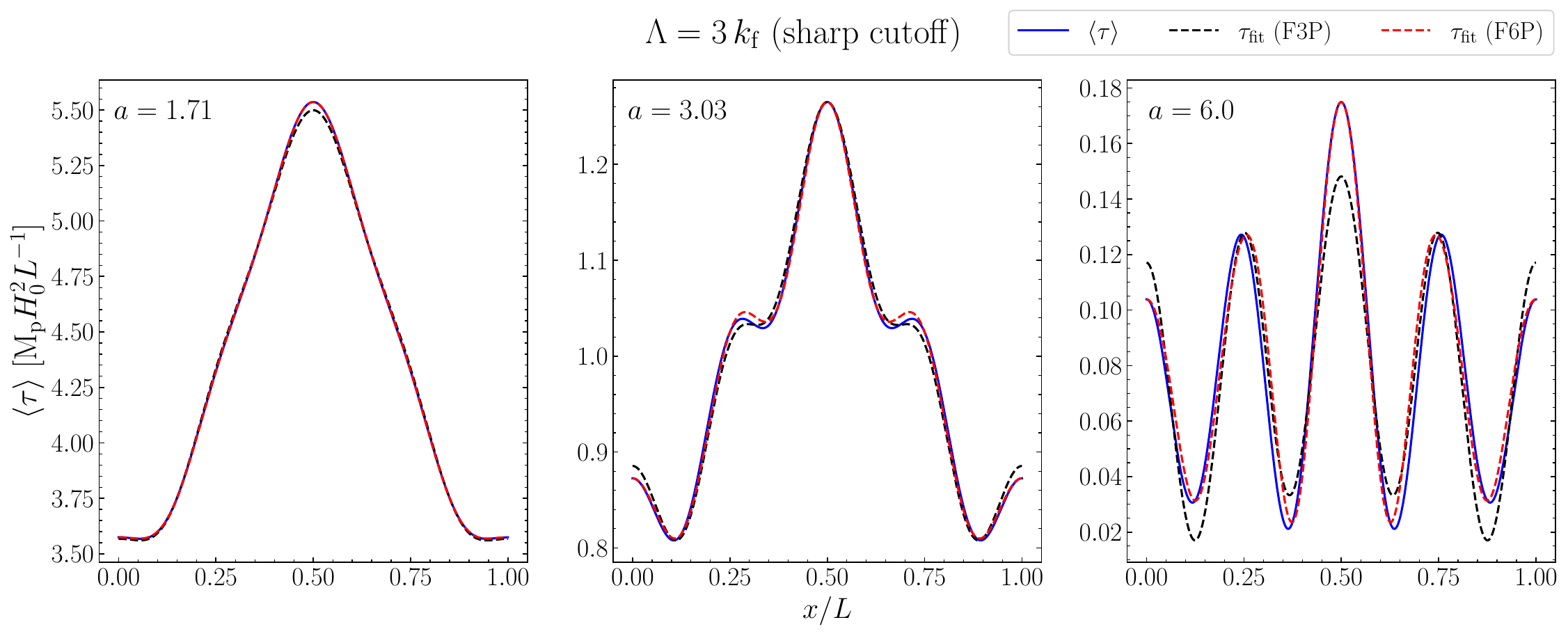}
        \end{subfigure}
    
        \begin{subfigure}{\textwidth}
            \centering
            \includegraphics[width=\textwidth]{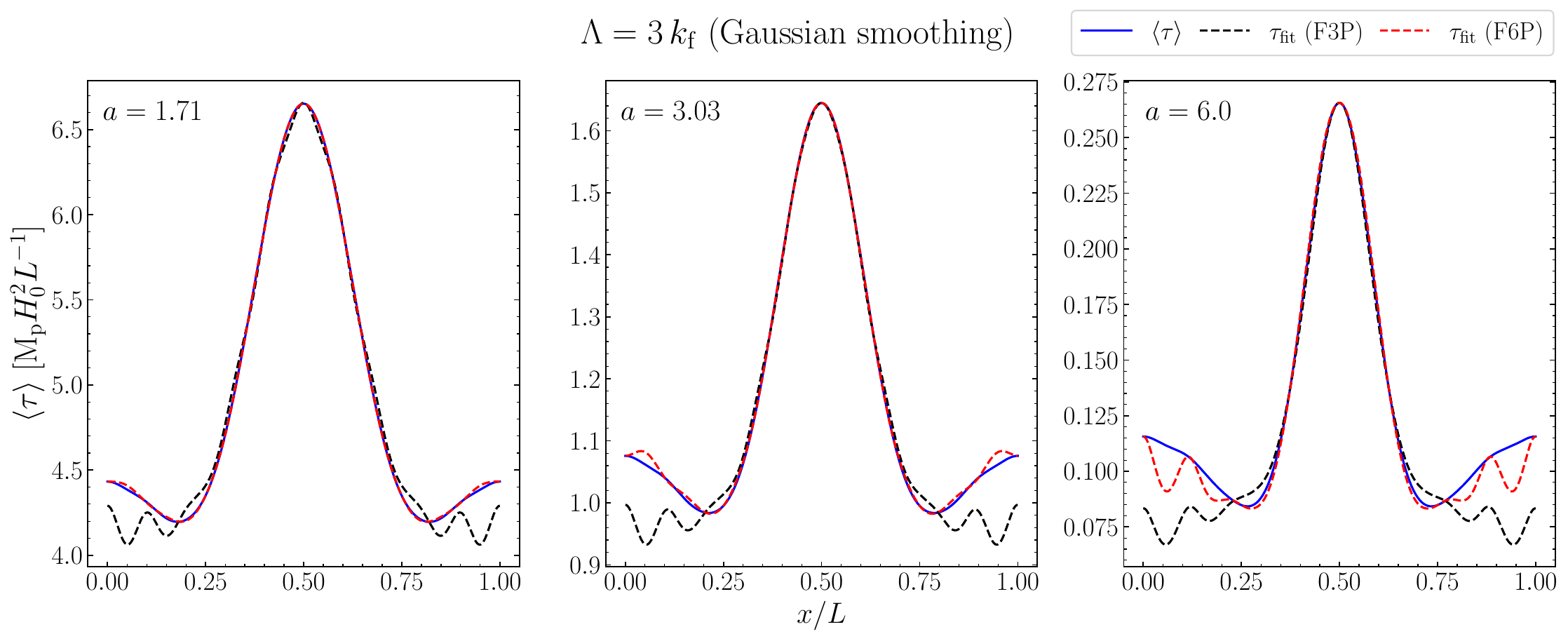}
        \end{subfigure}
        \caption{Fits to the averaged effective stress (using Eq.~\ref{eq:eft_fit_gen}, F3P, or its extension to quadratic order, F6P), with a sharp cutoff at $\Lambda = 3\,k_{\mathrm{f}}$. The solid blue line is the stress estimated from the simulations, and the dashed lines represent the fits using the F3P (black) and the F6P (red) models.}
        \label{fig:tau_fit}
    \end{figure}

Starting from the $\delta_{\ell}$, $\theta_{\ell}$, and $\langle\tau\rangle$ estimated by averaging at fixed $x$ over the eight realisations, we fit the coefficients of the expression
    \begin{equation}\label{eq:eft_fit_gen}
        \langle\tau\rangle = C_{0} + C_{1} \,\delta_{\ell} + C_{2} \,\theta_{\ell}
    \end{equation}
using the least-squares method.
The fit parameters $C_i$ are related to the expansion in Eq.~(\ref{eq:tau_expansion}). We refer to the resulting values for $c^2_\mathrm{s}$ and $c^2_\mathrm{v}$ as the F3P estimates (short for `fit using three parameters'). To check the robustness of the results, we also fit $\langle \tau\rangle$ using all possible terms up to second-order in $\delta_{\ell}$ and $\theta_{\ell}$, i.e. by adding $C_{3} \,\delta_{\ell}^{2}+ C_{4} \,\theta_{\ell}^{2} + C_{5} \,\delta_{\ell} \theta_{\ell}$ to the right-hand side of Eq.~(\ref{eq:eft_fit_gen}). We refer to the estimates for $c^2_\mathrm{s}$ and $c^2_\mathrm{v}$ obtained this way using the label F6P (short for `fit using six parameters'). Figure~\ref{fig:tau_fit} shows the accuracy of the fits (which are indicated by $\tau_{\mathrm{fit}}$) using $\Lambda = 3\,k_{\mathrm{f}}$. For the sharp filter (top row), both the three- and six-parameter fits match the estimated $\langle\tau\rangle$ well. At $a=6$, however, F6P better reproduces the peaks and valleys of the high-frequency oscillations seen in $\langle \tau \rangle$. When using Gaussian smoothing (bottom row), both F3P and F6P accurately reproduce $\langle \tau \rangle$ around the centre of the simulation volume, while F6P provides a better match to the measured stress near the edges of the box at all times.
    
\subsubsection{Comparing the different estimates} \label{sec:results_ctot2}
\begin{figure}
        \centering
        \includegraphics[width=0.75\textwidth]{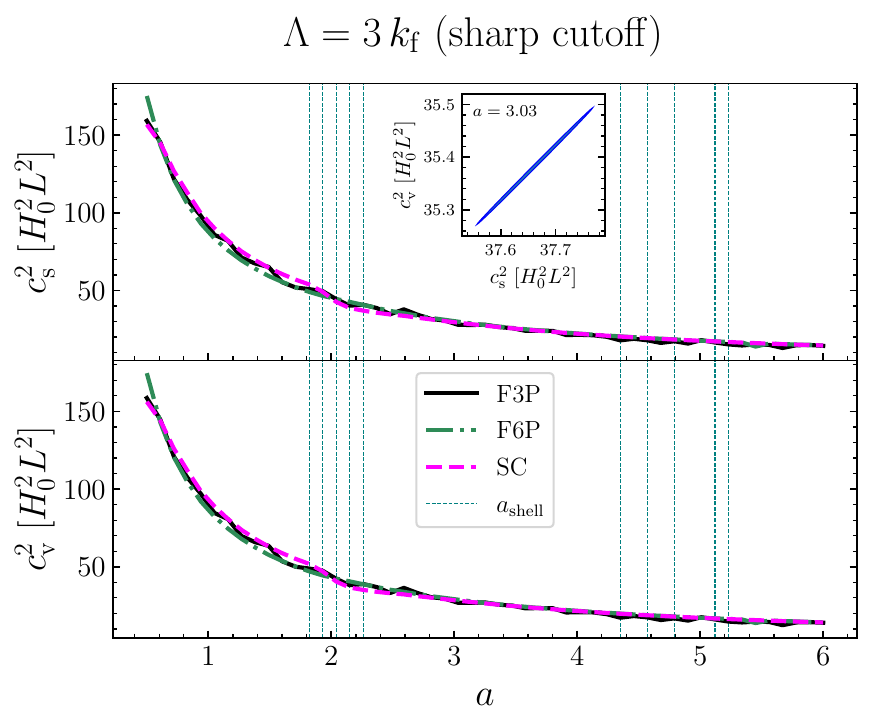}
        \caption{The $c_{\mathrm{s}}^{2}$ and $c_{\mathrm{v}}^{2}$ values obtained with the F3P, F6P, and SC estimators using a sharp filter with $\Lambda=3\,k_\mathrm{f}$. The dashed vertical lines indicate the times when a shell crossing is detected in \texttt{sim\_1\_11}. \emph{Inset}: The 68\% confidence ellipse for $c_{\mathrm{s}}^{2}$ and $c_{\mathrm{v}}^{2}$ at $a=3.03$ for the F3P estimator.}
        \label{fig:cs2_sharp}
    \end{figure}

At this point, it is instructive to compare the $c_{\mathrm{s}}^{2}$ and $c_{\mathrm{v}}^{2}$ values obtained from the two classes of estimators we have discussed. In Figure~\ref{fig:cs2_sharp}, we show the results obtained applying the F3P (solid black line), F6P (dash-dotted green line) and SC (dashed magenta line) methods to \texttt{sim\_1\_11} for the sharp-cutoff case. The three estimates are in excellent agreement with each other, thus suggesting that we can robustly measure the IR-UV coupling coefficients of the effective theory.

The best-fit values for $c_{\mathrm{s}}^{2}$ and $c_{\mathrm{v}}^{2}$ are similar to each other. Thus, their difference $c_{\mathrm{tot}}^{2}$ (the only combination entering the corrections to the matter power spectrum, see Section~\ref{sec:EFT_corrections}) is much smaller than the individual values. In the inset of Figure~\ref{fig:cs2_sharp}, we show the error ellipse for $c_{\mathrm{s}}^{2}$ and $c_{\mathrm{v}}^{2}$ obtained with the F3P method at $a=3.03$. The statistical uncertainties on the individual fit parameters are approximately 10\% but their difference is much more tightly constrained since the coefficients are highly correlated, as expected from the fact that $\delta_{\ell} \approx -\theta_{\ell}$.

\begin{figure}
        \centering
        \includegraphics[scale=0.5]{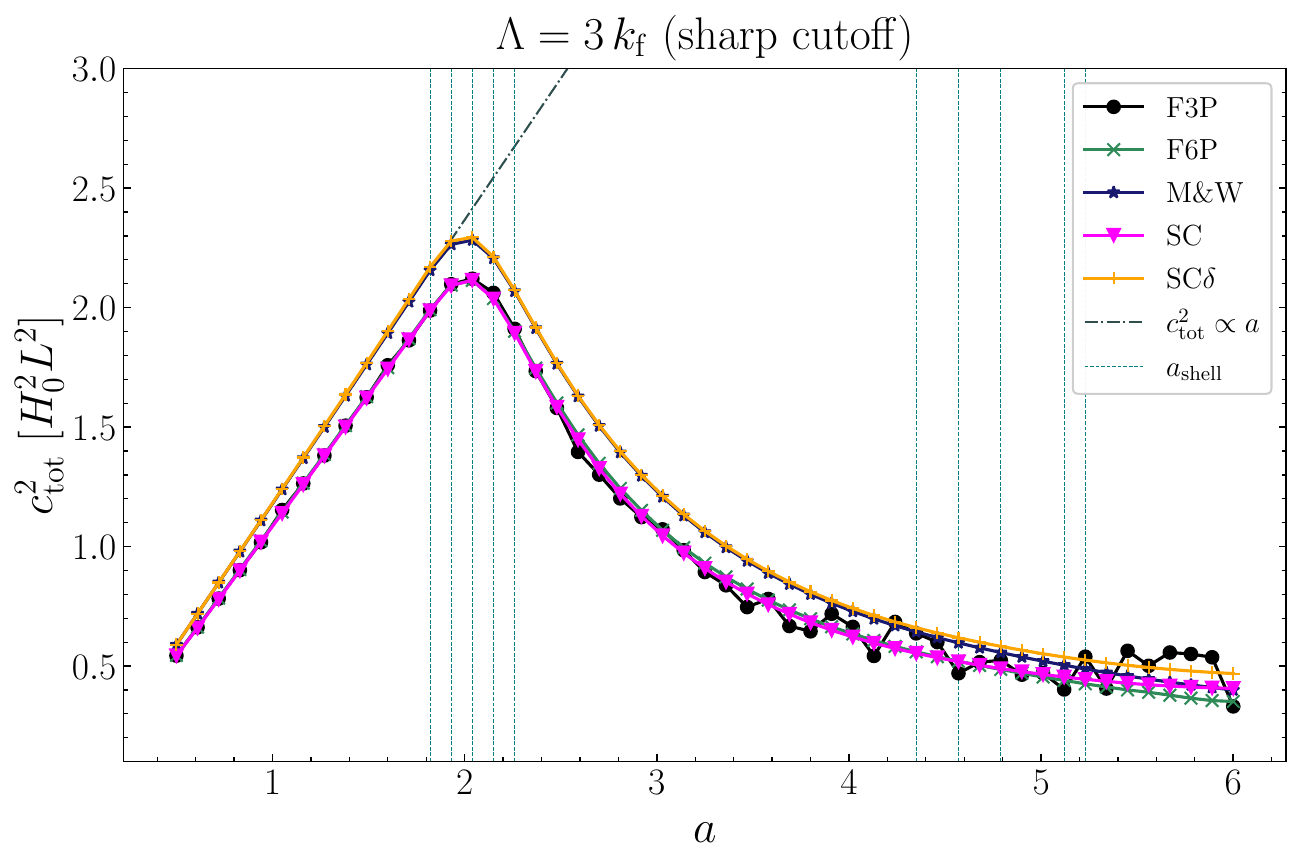}
        \caption{Variation of $c_{\mathrm{tot}}^{2}$ with $a$ for \texttt{sim\_1\_11} using the top-down estimators. The dash-dotted line corresponds to the scaling $c^{2}_{\mathrm{tot}} \propto a$
        generally assumed in applications of the EFTofLSS. The dashed vertical lines are as in Figure~\ref{fig:cs2_sharp}.}
        \label{fig:ctot2_ev_sharp}
    \end{figure}

In Figure~\ref{fig:ctot2_ev_sharp}, we show $c^{2}_{\mathrm{tot}}$ for \texttt{sim\_1\_11} as a function of time for the top-down estimators in case of a sharp cutoff with $\Lambda = 3\,k_{\mathrm{f}}$ (for most of our following results, we will relegate the plots for Gaussian smoothing to Appendix \ref{sec:gauss_appendix}). The dash-dotted green line represents the scaling $c^{2}_{\mathrm{tot}} \propto a$, which is the time-dependence assumed in the EFTofLSS literature based on perturbative considerations \cite[e.g.][]{carrasco_2012, hertzberg2014effective, carrasco2014, mw_2016}. All estimates satisfy this relation when the flow is in the single-stream regime. As the first set of shell crossings occur, however, the linear growth stops and $c^{2}_{\mathrm{tot}}$ steadily decreases with time until $a=6$. Because of this, $c^{2}_{\mathrm{tot}}$ assumes its peak value at $a = 2.04$. It has been argued that virialised structures should decouple from the long-wavelength dynamics and stop contributing to $c^2_\mathrm{tot}$ \cite{baumann_2012}. The decrease we see in $c^2_\mathrm{tot}$ might then reflect the fact that the 1D halos in \texttt{sim\_1\_11} are approaching or have reached virialisation. However, $c^{2}_{\mathrm{tot}}$ does not vanish because only half of the particles lie in multi-stream regions by $a=6$. These conjectures suggest that $c^2_\mathrm{tot}$ is sourced by perturbations that are entering the non-linear regime. In a sense, our toy model allows us to identify the response of the effective coupling constant to a single scale becoming non-linear. It is thus reasonable to expect that, in the presence of a continuous spectrum of perturbations, the time evolution of $c^2_\mathrm{tot}$ mirrors the flow of power into the non-linear regime. Lastly, we note that the M\&W and SC$\delta$ estimators give slightly larger values for $c^{2}_{\mathrm{tot}}$. This overestimation likely comes from assuming that $\delta_{\ell} = -\theta_{\ell}$.

\section{Performance of the EFTofLSS} \label{sec:results}
\subsection{Next-to-leading-order power spectrum in SPT and EFTofLSS}\label{sec:EFT_corrections}
SPT solves the system of equations (\ref{eq:spt_continuity_3D}--\ref{eq:spt_poisson_3D}) perturbatively. It is convenient to work in Fourier space, where spatial derivatives transform to products by the wavenumber. In the 1D case, one obtains
    \begin{equation}\label{eq:field_spt}
        \widetilde{\delta}(k, a) = a\,\widetilde{\delta}_{1}(k) + a^{2}\,\widetilde{\delta}_{2}(k) + a^{3}\,\widetilde{\delta}_{3}(k) 
        + \ldots \;,
    \end{equation}
with
    \begin{equation}\label{eq:SPTexp}
        \widetilde{\delta}_{n}(k)=
        \int F_n(k_1,\dots,k_n)\,\widetilde{\delta}_{1}(k_1)\,\dots\,
        \widetilde{\delta}_{1}(k_n)\,\delta_\mathrm{D}\left(k-\sum_{i=1}^n k_i\right)\,\dd k_1 \dots \dd k_n\;,
    \end{equation}
where $a\,\widetilde{\delta}_{1}(k)$ represents the solution of the linearised system of equations and the $n^\mathrm{th}$-order kernel $F_n(k_1,\dots,k_n)$ quantifies the coupling between Fourier modes due to the non-linear dynamics. The drawback of SPT is that the integrals in Eq.~(\ref{eq:SPTexp}) receive contributions from Fourier modes with $k>k_\mathrm{NL}$ for which the SPT equations do not apply. In our simplified setup with periodic boundary conditions, and which only contains linear modes with $k/k_\mathrm{f}=\pm 1$ and $\pm 11$, $\widetilde{\delta}_{2}(k)$ is non-zero only for $k/k_\mathrm{f}=\pm 2,\pm 10, \pm 12, \pm 22$, while $\widetilde{\delta}_{3}(k)$ is non-zero for $k/k_\mathrm{f}=\pm 1, \pm 3, \pm 9, \pm 11, \pm 13, \pm 21, \pm 22, \pm 23, \pm 33$. Therefore, the SPT prediction for the matter power spectrum $P(k)=\widetilde{\delta}(k)\,\widetilde{\delta}(-k)$ evaluated at $k=k_\mathrm{f}$ to next-to-leading order (NTLO) is $P_\mathrm{SPT}(k_\mathrm{f})=P_{11}(k_\mathrm{f})+2\,P_{13}(k_\mathrm{f})$ where $P_{ij}(k)=a^{i+j}\,\widetilde{\delta}_i(k)\,\widetilde{\delta}_j(-k)$. 

\begin{figure}
        \centering
        \includegraphics[width=\textwidth]{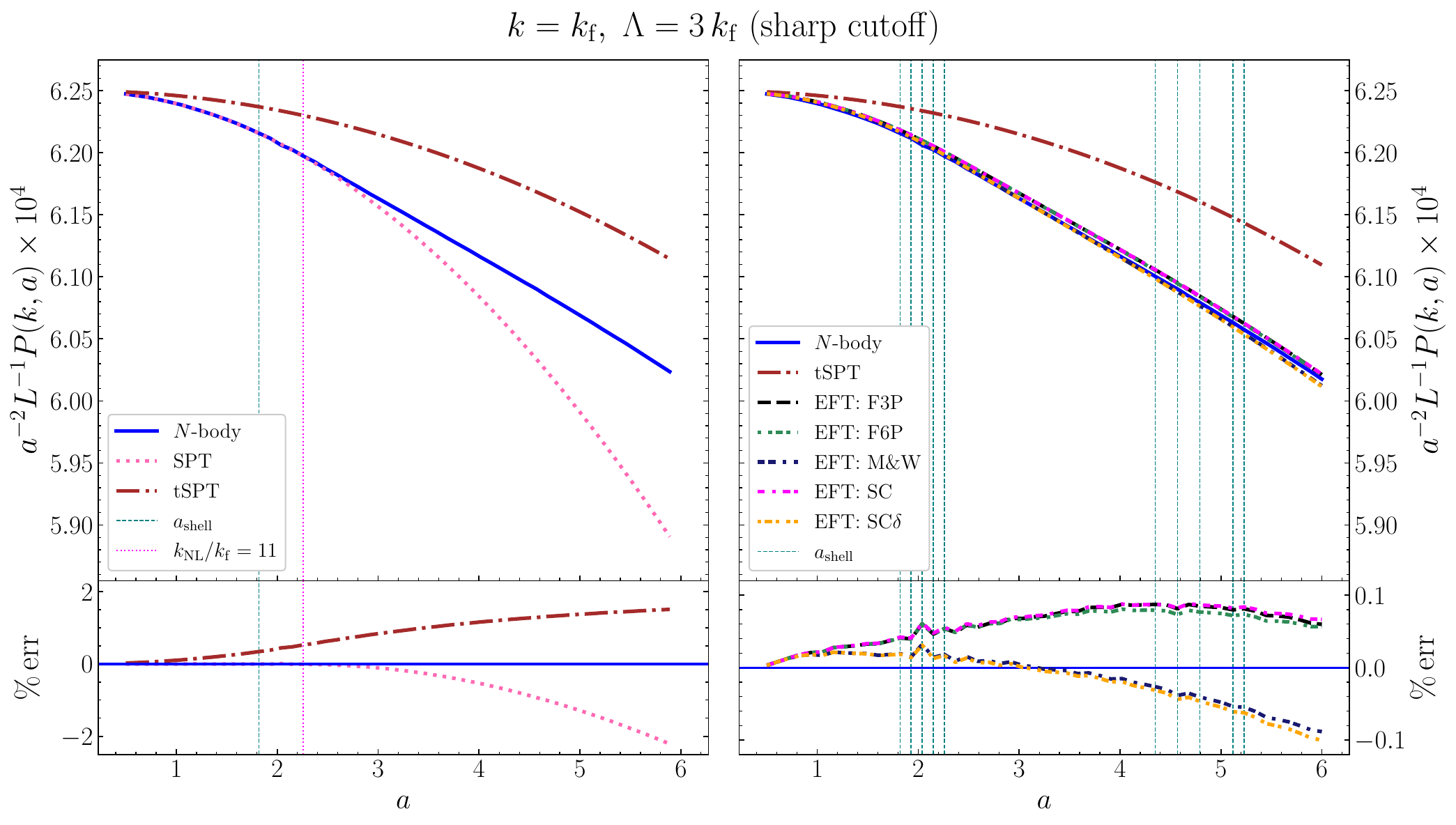}
        \caption{The solid blue line appearing in both panels shows the time evolution of the matter power spectrum in \texttt{sim\_1\_11} evaluated at $k=k_\mathrm{f}$. The matter density has been smoothed using a sharp filter with $\Lambda=3\,k_\mathrm{f}$. In the left panel, the $N$-body result is compared to the NTLO predictions from SPT (dotted pink line) and tSPT (dash-dotted brown line). Two important timescales are indicated with vertical lines: the time of the first shell crossing (dashed teal line) and the time at which $k_{\mathrm{NL}} / k_{\mathrm{f}}=11$ (dotted magenta line). In the right panel, the $N$-body result is contrasted with the NTLO predictions from the EFTofLSS obtained using five different top-down estimators for $c^2_\mathrm{tot}$. The dashed vertical lines are as in Figure~\ref{fig:cs2_sharp}. The bottom panels show the percentage deviation of the models from the simulation. Note that the vertical scale differs in the left and right panels. In the bottom right panel, we omit tSPT to better visualise the EFT errors.}
        \label{fig:spt_spec}
    \end{figure}

In the left panel of Figure~\ref{fig:spt_spec}, we compare this result (dotted pink line) to the spectrum measured from \texttt{sim\_1\_11} (solid blue line). In order to make non-linearities more visible, we erase the linear evolution by plotting $a^{-2}\,P(k_\mathrm{f})$ vs $a$. It has been shown that in 1D, the $n$-loop SPT prediction converges to the Zel'dovich solution before shell-crossing in the limit $n\to \infty$ \cite{mw_2016}. It is therefore not surprising that the SPT prediction in our setup asymptotically matches the $N$-body spectrum in the small-$a$ limit. Perhaps more interestingly, the prediction remains extremely accurate even beyond the first shell crossing, and up to $a\approx3$ (when nearly $30\%$ of the matter lies in multi-stream regions). At this epoch, $k_\mathrm{NL}/k_\mathrm{f}\simeq 9$ and the modes with $k=\pm11\,k_\mathrm{f}$ are non perturbative. At $a = 6$, the SPT model underestimates the $N$-body result by $\approx 2.2\%$.

Next, we investigate the effect of introducing a cutoff in the SPT loop integrals. The dash-dotted brown line in Figure~\ref{fig:spt_spec} shows the NTLO SPT spectrum obtained using a sharp cutoff with $\Lambda = 3\,k_{\mathrm{f}}$. We refer to this model as truncated SPT (tSPT, in short). With our initial conditions, this is equivalent to only accounting for the mode coupling between the
two linear modes with $k=\pm k_\mathrm{f}$. Therefore, the difference between the SPT and tSPT models shows the contribution of the linear modes with $k=\pm11\,k_\mathrm{f}$ to $P_\mathrm{SPT}(k_\mathrm{f})$. Their net effect is to slow down the growth of the long-wavelength perturbation. The tSPT model overpredicts $P(k_\mathrm{f})$ at all times. Early on, when $k_\mathrm{NL}< 11\, k_\mathrm{f}$, and the linear short-wavelength modes in our setup are perturbative, SPT is much more accurate than tSPT. The error in SPT remains smaller than in tSPT until $a\simeq 5$, at which time they cross over. The tSPT error at $a=6$ is $\approx 1.5\%$.

In the EFTofLSS, the expansion of Eq.~(\ref{eq:tau_expansion}) is substituted into the 1D equivalents of Eqs.~(\ref{eq:coarse_continuity_3D}) and (\ref{eq:coarse_euler_3D}) to obtain (along with the Poisson equation) the dynamical equations for the effective fluid
    \begin{align}
        \dot{\delta}_{\ell} + \frac{1}{a}\partial_{x}\,[(1 + \delta_{\ell})\,U] &= 0 \label{eq:coarse_continuity}\;, \\
        \dot{U} + \frac{\dot{a}}{a}\,U + \frac{1}{a}\,U\,\partial_{x}U &=- \frac{1}{a}\,\partial_{x}\phi_{\ell} - \frac{c^{2}_{\mathrm{s}}}{a}\,\partial_{x}\delta_{\ell} - \frac{c^{2}_{\mathrm{v}}}{a}\,\partial_{x}\theta_{\ell}-
        \frac{1}{a\rho_\ell}\,\partial_x J
        \label{eq:coarse_euler}\;.
    \end{align}
At linear order in the density perturbations, 
    \begin{equation}
        \left(\partial_t^2+2\,\frac{\dot{a}}{a}\,\partial_t-4\pi G\rho_\mathrm{b}\right) \delta_{1\ell} =
        \frac{1}{a^2}\,c^2_\mathrm{tot}\,\partial_x^2\delta_{1\ell}
        +\frac{1}{a^2\,\rho_\ell}\,\partial^2_x J \;,
    \end{equation}
which differs from the corresponding tSPT equation because the terms on the right side do not vanish\footnote{References \cite{garny2023perturbation, garny2023perturbation2} have recently introduced a formalism which also includes higher-order cumulants in the perturbative expansion.}. Neglecting for a moment the contribution from the stochastic term which will be discussed separately in Section~\ref{sec:stochastic}, the leading-order correction to the tSPT solution is
    \begin{equation}\label{eq:d3_corr}
        \widetilde{\delta}^\mathrm{c}_{\ell}(k, a) = 
        \alpha_\mathrm{c}\,a\,k^{2}\, \widetilde{\delta}_{1\ell}(k)\;,
    \end{equation}
where
 \begin{equation}\label{eq:alpha_c_def}
        \alpha_{c} = \frac{1}{a}\,\int_{0}^{\infty} G(a, a') \,c^{2}_{\mathrm{tot}}(a')\, a'\, \dd{a'}\;,
    \end{equation}
and, in the EdS universe, the retarded Green's function assumes the form \cite[e.g.][]{pz_2013, hertzberg2014effective}
 \begin{equation}\label{eq:greens_fun}
        G(a, a') = \Theta(a-a')\,\frac{2}{5H^{2}_{0}}\,\left[\left(\frac{a'}{a}\right)^{3/2} - \frac{a}{a'}\right]\;.
    \end{equation}
It follows that the EFTofLSS power spectrum to NTLO is   
    \begin{equation}\label{eq:PS_corr}
        P_{\mathrm{EFT}} = P_{\mathrm{tSPT}} + 2 \,\alpha_{c}\,k^{2}\,\widetilde{W}_\Lambda^2\,P_{11}\;.
    \end{equation}
We obtain top-down estimates for $\alpha_c$ by numerically integrating Eq.~(\ref{eq:alpha_c_def}) and using the functions $c_{\mathrm{tot}}^{2}(a)$ measured from the simulations and presented in Figure~\ref{fig:ctot2_ev_sharp}. Since our simulations start at $a=0.5$, we assume $c_{\mathrm{tot}}^{2} \propto a$ for $a<0.5$. As shown in the right panel of Figure~\ref{fig:spt_spec}, the resulting power spectra provide an excellent match to the $N$-body results for $k=k_\mathrm{f}$. Note, however, that the spectra are never asymptotically correct in any regime, and it is not clear how this could be rectified at higher order. Nonetheless, deviations are always smaller than $0.06\%$ even at $a=6$ when $\Lambda=3\,k_\mathrm{f}$ is only 1.5 times larger than $k_\mathrm{NL}$.
This result provides evidence that the effective dynamics accurately models the feedback of the short modes onto the growth rate of the long ones. All the top-down estimators for $c^2_\mathrm{tot}$ we used essentially give the same result with the sharp filter while some dissimilarities are noticeable for Gaussian smoothing (see Appendix \ref{sec:gauss_appendix}).

\begin{figure}
        \centering
        \includegraphics[scale=0.5]{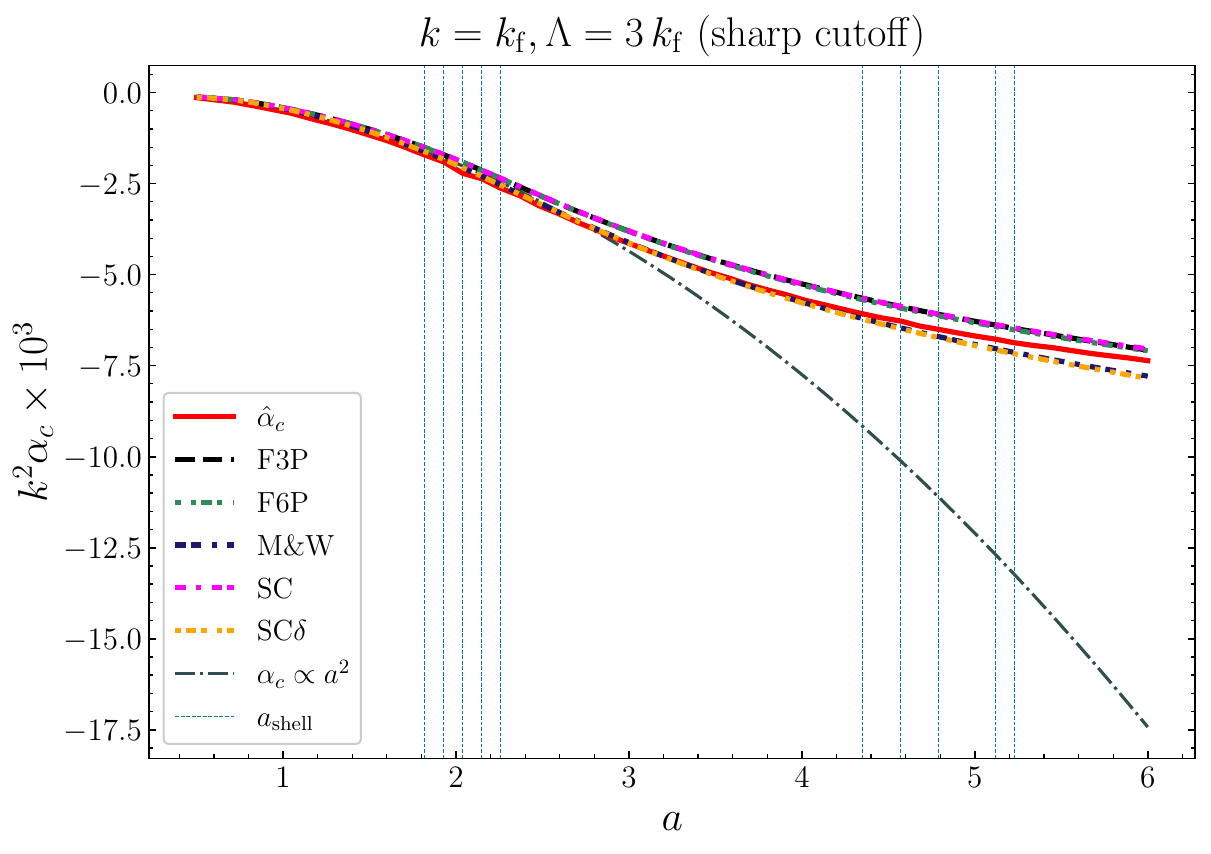}
        \caption{Comparison of $\alpha_{c}$ between the bottom-up estimator (solid green line) and the top-down estimators (dashed lines).
        The dashed vertical lines are as in Figure~\ref{fig:cs2_sharp}.}
        \label{fig:alpha_c_sharp}
    \end{figure}

Assuming -- as done in the literature \cite[e.g.][]{carrasco_2012, carrasco2014, hertzberg2014effective, mw_2016} -- that $c_{\mathrm{tot}}^{2} \propto a$ for all $a$, we can write $c_{\mathrm{tot}}^{2} = c^{2}_{0}\,a$ (where $c^{2}_{0}$ is the estimated value at $a=1$). Using this expression, we can analytically integrate Eq. (\ref{eq:alpha_c_def}), which gives the relation
    \begin{equation}\label{alpha_c_an}
        \alpha_{c} = -\frac{c^{2}_{0}a^{2}}{9H_{0}^{2}} \;.
    \end{equation}   
In our setup, using this scaling gives inaccurate results for $a>3$. This is shown in Figure~\ref{fig:alpha_c_sharp} where we plot the time evolution of the EFT correction to the NTLO power spectrum for $k=k_\mathrm{f}$ and $\Lambda=3\,k_\mathrm{f}$. The solid line represents the value of $\alpha_{c}$ that perfectly reproduces the $N$-body power spectra, i.e.
    \begin{equation}\label{eq:alpha_c}
        \hat{\alpha}_{c}= \frac{P_{N\mathrm{-body}} - P_{\mathrm{tSPT}}}{2\,k^{2}\,\widetilde{W}^{2}_\Lambda\,P_{11}} \;.
    \end{equation}
This is the `bottom-up' approach predominantly used in the EFTofLSS literature. As expected from the right panel in Figure~\ref{fig:spt_spec}, the top-down estimates for $\alpha_{c}$ cluster around the bottom-up value. However, the approximation in Eq.~(\ref{alpha_c_an}) (dark grey dash-dotted line) departs from $\hat{\alpha}_{c}$ at late times. This happens because the actual value of $c^2_\mathrm{tot}$ scales proportionally to $a$ only before shell crossing takes place and decreases afterwards. As $\alpha_{c}$ evaluated at a given $a$ retains information from all $a'< a$ as we integrate over the past, the scaling $\alpha_c\propto a^2$ continues to hold until $a\approx 3$. However, at later times, $\alpha_{c}$ grows slower than the $a^{2}$ scaling, consistent with a decreasing value of $c_{\mathrm{tot}}^{2}$.

\subsection{Stochastic term}\label{sec:stochastic}
In our discussion so far, we have ignored the stochastic part of the effective stress as its impact on the matter power spectrum is expected to be suppressed at $k<\Lambda$ with respect to the deterministic NTLO corrections \cite{baumann_2012}. The reason is twofold. First, the variance of the stochastic fluctuations in a patch of size $\Lambda$ is determined by the properties of the short-wavelength fluctuations with $k>\Lambda$ and should be dominated by the perturbations that are turning non-linear with $k\simeq k_\mathrm{NL}$. Second, a region of size $k^{-1}$ contains many sub-domains of size $\Lambda^{-1}$ and thus the amplitude of statistical fluctuations is suppressed by averaging over the large number of sampled sub-regions. In this section, we investigate the properties of the stochastic contribution in our simulations. We compute $J$ as the difference between the effective stress measured from \texttt{sim\_1\_11} and that reconstructed from the $c_{\mathrm{s}}^{2}$ and $c_{\mathrm{v}}^{2}$ using the SC and F3P estimators. For $a<3$ we find that $J$ only contributes to $\tau$ at the sub-percent level, while the stochastic fraction reaches $40\%$ at $a=6$ with the sharp filter. However, most of this contribution comes from higher-$k$ modes, and does not enter the $k = k_{\mathrm{f}}$ calculations. Also, note that our measurement of $J$ is degenerate to higher-order terms in the expansion of $\langle \tau\rangle$. Including the stochastic term in Eq.~(\ref{eq:coarse_euler}), gives a correction to the tSPT overdensity \cite[e.g.][]{pz_2013}
    \begin{equation}\label{eq:stoch_green}
        \widetilde{\delta}_{\mathrm{J}}(k, a) = -\frac{k^{2}}{\rho_{b}}\int_{0}^{\infty}G(a, a')\,J(k, a')\, \dd{a'} \;,
    \end{equation}
where $G(a, a')$ is the Green's function defined in Eq.~(\ref{eq:greens_fun}). We find that this term makes negligible contributions to the matter power spectrum at $k=k_\mathrm{f}$. The auto-spectrum of ${\delta}_{\mathrm{J}}$ (hereafter, $P_{\mathrm{J}{\mathrm{J}}}$) is always of the order of a part per million, independently of the estimator used to determine $c^2_\mathrm{tot}$.

\subsection{Dependence on the separation of scales}
\begin{figure}
        \centering
        \includegraphics[width=\textwidth]{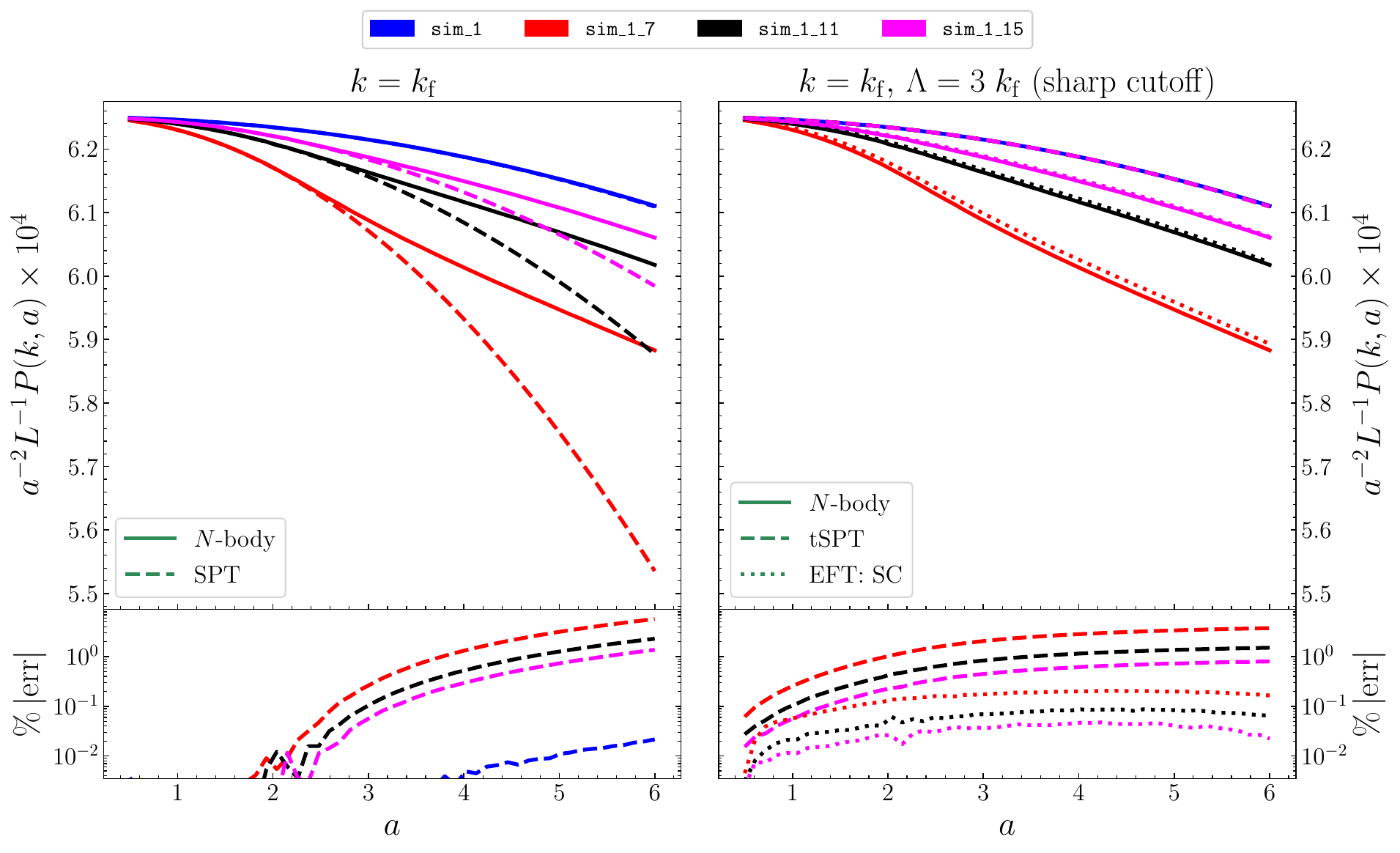}
        \caption{As in Figure~\ref{fig:spt_spec}, but for a suite of simulations with different separation between the wavelengths of the non-vanishing linear perturbations. Note that: (i) all the tSPT lines coincide in the top-right panel; (ii) the colors of the lines have different meanings from  Figure~\ref{fig:spt_spec}; (iii) contrary to Figure~\ref{fig:spt_spec}, we plot the absolute value of the percentage deviation in the bottom panels.}
        \label{fig:spec_comp}
    \end{figure}

An effective theory assumes that the scales it describes must be well-separated from the UV scales. In our simple setup, we can change the separation between wavelengths and see how the different flavours of perturbation theory compare with the numerical solution as a consequence. To this end, we run two further simulations with the same amplitude ratio as \texttt{sim\_1\_11}, but with $n_{1} = 1, n_{2} = 7$ (hereafter \texttt{sim\_1\_7}), and $n_{1} = 1, n_{2} = 15$ (hereafter \texttt{sim\_1\_15}). We show the effect of this change in separation in the left panel of Figure~\ref{fig:spec_comp}, by plotting the non-linear evolution of the power in $k=k_{\mathrm{f}}$ obtained from the simulations (solid lines). We also show the SPT power spectra in each case (dashed lines). For \texttt{sim\_1} (the simulation run with only the long-wavelength mode), SPT matches the $N$-body spectrum well at all times, as there is no linear high-amplitude mode. Comparing \texttt{sim\_1} and the other simulations, we see that presence of the linear high-amplitude, short-wavelength mode dampens the $N$-body spectrum at $k=k_{\mathrm{f}}$. The strength of this dampening increases with decreasing separation, as expected. For all simulations, the SPT and  $N$-body spectra agree well until $a \approx 3$. At later times, however, SPT fails to correctly capture the non-linear evolution and underestimates the power by a few per cent. The absolute value of the error (shown in the bottom panel) decreases with increasing the separation between the linear modes. In the right panel, we plot the tSPT spectra (dashed lines) in addition to the $N$-body results, when using a sharp cutoff with $\Lambda=3\,k_\mathrm{f}$. All the tSPT spectra are identical to each other as truncation removes the high-amplitude mode. It follows that tSPT overestimates the power by several percent and the error with respect to the simulations increases with decreasing separation between the linear modes. Finally, we plot the EFT spectra (dotted lines) obtained with the SC estimator for $c^2_\mathrm{tot}$ (all the estimators are in good agreement for all simulations). These predictions remain accurate across simulations to $\approx 0.1\%$ at all times. As expected, the accuracy is higher for the initial conditions with the largest separation between the modes. These results show that the EFT corrections are robust to change in separation between the linear modes.

\subsection{Considerations about renormalisation}
In general, the tSPT loop integrals $P_{13}$ and $P_{22}$ depend on $\Lambda$ due to truncation of the linear power spectrum which delimits the range of integration. Similarly, the EFT correction in Eq.~(\ref{eq:PS_corr}) depends on $\Lambda$ through $\alpha_c$ (and thus $c^2_\mathrm{tot}$) which is determined from the smoothed perturbations. Since $P_{13}$ and the EFT correction have the same $k$ dependence as $k\to 0$, references \cite{carrasco_2012, hertzberg2014effective, carrasco2014} proposed to renormalise the coupling coefficient between long- and short-wavelength perturbations to get $\Lambda$-independent predictions. This mirrors the renormalisation procedure used in field theory to control UV divergences. For the moment, we only consider the renormalisation of the $P_{13}$ contribution. We later remark on a similar scheme for $P_{22}$ and its applicability in our setup. We begin by choosing a `renormalisation scale' $\mu < \Lambda$ below which one can trust the validity of SPT.
The `bare' effective parameter $\alpha_c$ is then decomposed into a ($\Lambda$-independent) renormalised part and a ($\Lambda$-dependent) counterterm
    \begin{equation}
        \alpha_c(\Lambda, \mu, a) = \alpha_\mathrm{ren}(\mu, a) + \alpha_\mathrm{ctr}(\Lambda, \mu, a)\;,
        \label{eq:lambdarenorm}
    \end{equation}
so that $\alpha_\mathrm{ren}(\mu, a)$ accounts for the actual correction to the dynamics generated by the non-perturbative scales while $\alpha_\mathrm{ctr}(\Lambda, \mu, a)$ removes the spurious contributions introduced by using SPT beyond its range of validity within the loop integrals. The renormalised parameter $\alpha_{\mathrm{ren}}$ is defined by modifying the bare parameter from the cutoff scale $\Lambda$ to the renormalisation scale $\mu$, in such a way that the physical observable (in our case the EFT power spectrum) is unaltered\footnote{As in field theory, this is governed by the renormalisation group equations. For a review of these techniques, see e.g. \cite{morris1994exact}. For an example of its implementation for the EFTofLSS, see \cite{carroll2014}.}. In practice, this requires that the counterterm exactly cancels the difference between the full $\Lambda$-dependent $P_{13}$ contribution and the renormalised contribution which depends on $\mu$.
    \begin{equation}
        2\,\alpha_{\mathrm{ctr}}(\Lambda, \mu, a)\,k^{2}\,\widetilde{W}^2_{\Lambda}(k)\,P_{11}(k, a)=-2\,\left[P_{13}^\mathrm{tSPT}(k,\Lambda, a)-
        P_{13}^\mathrm{tSPT}(k,\mu, a)
        \right]\;.
        \label{eq:rencond}
    \end{equation}
Since, in the limit\footnote{Note that $\widetilde{W}_\Lambda(k) \to 1$ in the limit $k \to 0$ for any low-pass filter.} $k\to 0$,
     \begin{equation}
         P_{13}^\mathrm{tSPT}(k,\Lambda, a) \to -a^{2}f(\Lambda)\,k^2\,P_{11}(k,a) \;,
     \end{equation}
where
    \begin{equation}
         f(\Lambda)=\int \frac{P_{11}(k,a=1)}{k^2}\,\widetilde{W}^2_{\Lambda}(k)\,\mathrm{d}k
    \end{equation}
is the variance of the matter displacement field\footnote{For \texttt{sim\_1\_11}, $f(\Lambda)=A_{1}^{2}\,\widetilde{W}_{\Lambda}(k_{\mathrm{f}}) / 2\,k_{\mathrm{f}}^{2} + A_{2}^{2}\,\widetilde{W}_{\Lambda}(11\,k_{\mathrm{f}})/2\,(11\,k_{\mathrm{f}})^{2}$ where the $A_{i}$ are the amplitudes of the linear overdensity defined in Eq.~\eqref{eq:in_den}, and any $\Lambda$-dependence is contained in the smoothing kernel $\widetilde{W}_{\Lambda}$.} at $a=1$, one, finally, gets
     \begin{equation}
         \alpha_{\mathrm{ctr}}(\Lambda, \mu, a)= -a^{2}\,[f(\Lambda)-f(\mu)]\;.
         \label{eq:guesscounter}
     \end{equation} 
Therefore, the full non-linear power spectrum for $k\to 0$ (ignoring the $P_{22}$ contribution) is
    \begin{equation}
        P_\mathrm{EFT}(k, a)=\left[1+2\,\alpha_{\mathrm{ren}}(\mu, a)\,k^{2}\right]\,P_{11}(k, a)+2\,P_{13}^\mathrm{tSPT}(k,\mu, a)\;.
        \label{eq:renormspectrun}
    \end{equation}
The final expression for the power spectrum is thus $\Lambda$-independent.

\begin{figure}
    \centering
    \includegraphics[scale=0.5]{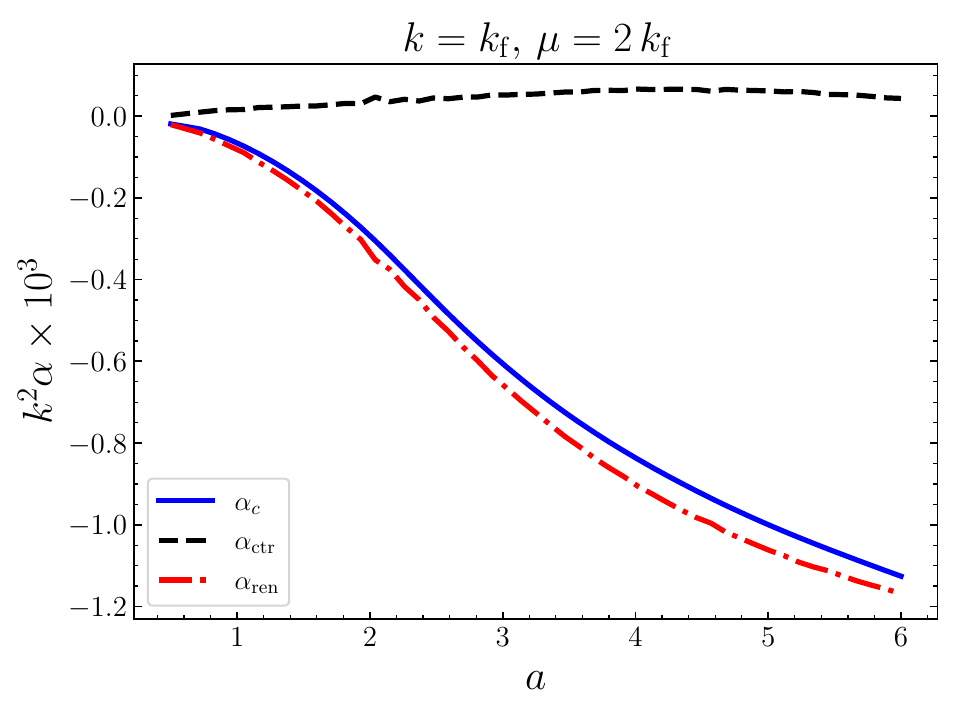}
    \caption{Time evolution of the bare coefficient $\alpha_{c}$ (solid blue line), the counterterm $\alpha_{\mathrm{ctr}}$ (dashed black line), and the renormalised coefficient $\alpha_{\mathrm{ren}}$ (dash-dotted red line).}
    \label{fig:counterterm_dep}
\end{figure}

In order to determine the terms appearing in Eq.~(\ref{eq:lambdarenorm}), we proceed as follows. First, we use Eq.~(\ref{eq:renormspectrun}) with $k=k_\mathrm{f}$ to measure $\alpha_{\mathrm{ren}}$ by matching the power spectrum of \texttt{sim\_1\_11}. Subsequently, we determine the counterterm by taking the difference between the bare $\alpha_c$ (measured from the simulations with our top-down approach) and its renormalised counterpart. Figure~\ref{fig:counterterm_dep} compares the time evolution of $\alpha_c$ (solid), $\alpha_\mathrm{ctr}$ (dashed), and $\alpha_\mathrm{ren}$ (dash-dotted) obtained using $\mu=2\,k_\mathrm{f}$. Three things are worth noticing. First, the counterterm is always positive while the other two quantities are negative. Second, $\alpha_\mathrm{ctr}$ and $\alpha_\mathrm{ren}$ do not share the same time evolution (contrary to the guess sometimes made in the literature \cite[e.g.][]{carrasco_2012}). In fact, the measured $\alpha_\mathrm{ctr}$ deviates from the expected $a^{2}$ scaling, especially at late times when the EFT prediction becomes less accurate. Third, the feedback due to the UV modes on the IR modes is much smaller than the non-linear interaction between the modes with $|k|/k_{\mathrm{f}}=1$ (as evidenced by the relatively small value of $\alpha_\mathrm{ctr}$).

A similar renormalisation scheme has been proposed to cancel the $\Lambda$-dependence of $P_{22}^\mathrm{tSPT}$ by using the spectrum generated by the stochastic fluctuations \cite{carrasco_2012,hertzberg2014effective} under the assumption that $P_{\mathrm{J}\mathrm{J}}\propto k^4$ when $k\to 0$ in analogy with Peebles' argument for uncorrelated small-scale displacements
that conserve momentum \cite{peebles1974gravitational}. However, this idea is not applicable to our configuration as $P_{22}^\mathrm{tSPT}$ is non-zero only for a discrete set of wavenumbers while $P_{\mathrm{J}\mathrm{J}}$ gets contributions at all wavenumbers. For instance, $P_{22}^\mathrm{tSPT}(k_\mathrm{f},\Lambda, a)=0$,
while $P_{\mathrm{J}\mathrm{J}}(k_\mathrm{f}, a)$ is small but non zero. Therefore, no renormalisation is possible.

\subsection{Long-term evolution of the effective speed of sound}
\begin{figure}
        \centering
        \includegraphics[scale=0.55]{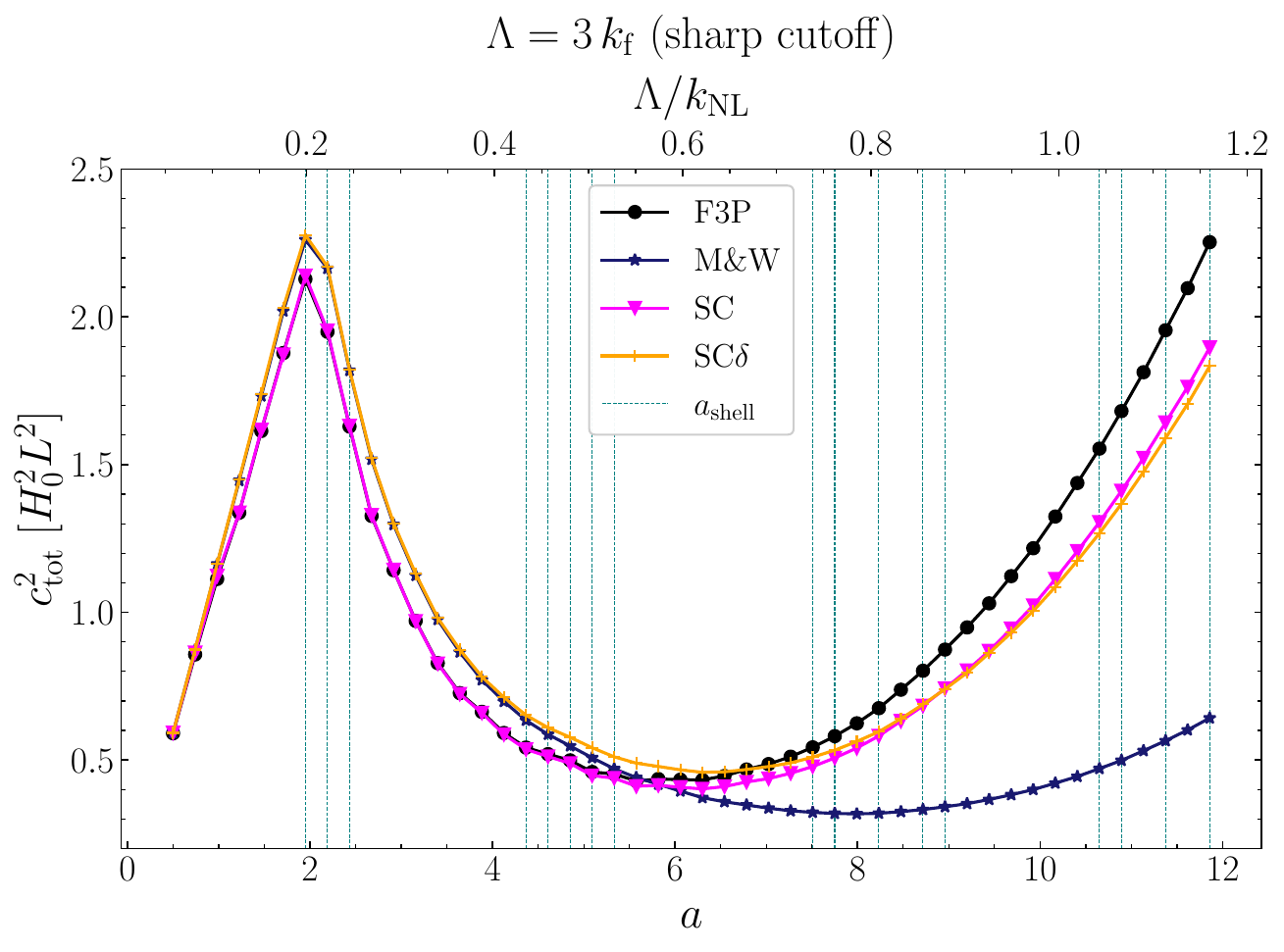}
        \caption{The long-term behaviour of $c_{\mathrm{tot}}^{2}$ for \texttt{sim\_1\_11}.}
        \label{fig:ctot2_ev_long}
    \end{figure}

Although our main simulations stop at $a=6$, we run a test simulation with the same setup as \texttt{sim\_1\_11} which extends to $a\approx 12$. The objective of this run is to understand how the EFT coefficients behave at very late times. In Figure \ref{fig:ctot2_ev_long} we show the evolution of $c_{\mathrm{tot}}^{2}$ with $a$. On the $x$-axis (top) we also show the ratio $\Lambda/k_{\mathrm{NL}}$. For $\Lambda = 3\,k_{\mathrm{f}}$, as this ratio approaches unity, the $3\,k_{\mathrm{f}}$ mode enters the non-linear regime. We note that after dropping to a small value after shell crossing $c_{\mathrm{tot}}^{2}$ starts increasing after $a=6$. At $a=6$, $\Lambda/k_{\mathrm{NL}} \approx 0.6$, indicating that $\Lambda$ and $k_{\mathrm{NL}}$ are not well-separated. At later times, the increase of $c_{\mathrm{tot}}^{2}$ may therefore be attributed to the $3\,k_{\mathrm{f}}$ mode entering the non-linear regime. By $a=12$, $\Lambda > k_{\mathrm{NL}}$, meaning that this epoch is beyond the regime of applicability of the EFTofLSS. We also note that the M\&W estimator deviates from the others after $a=6$. This is because the estimator assumes linear theory which is no longer valid in this regime.

\begin{figure}
        \centering
        \includegraphics[scale=0.55]{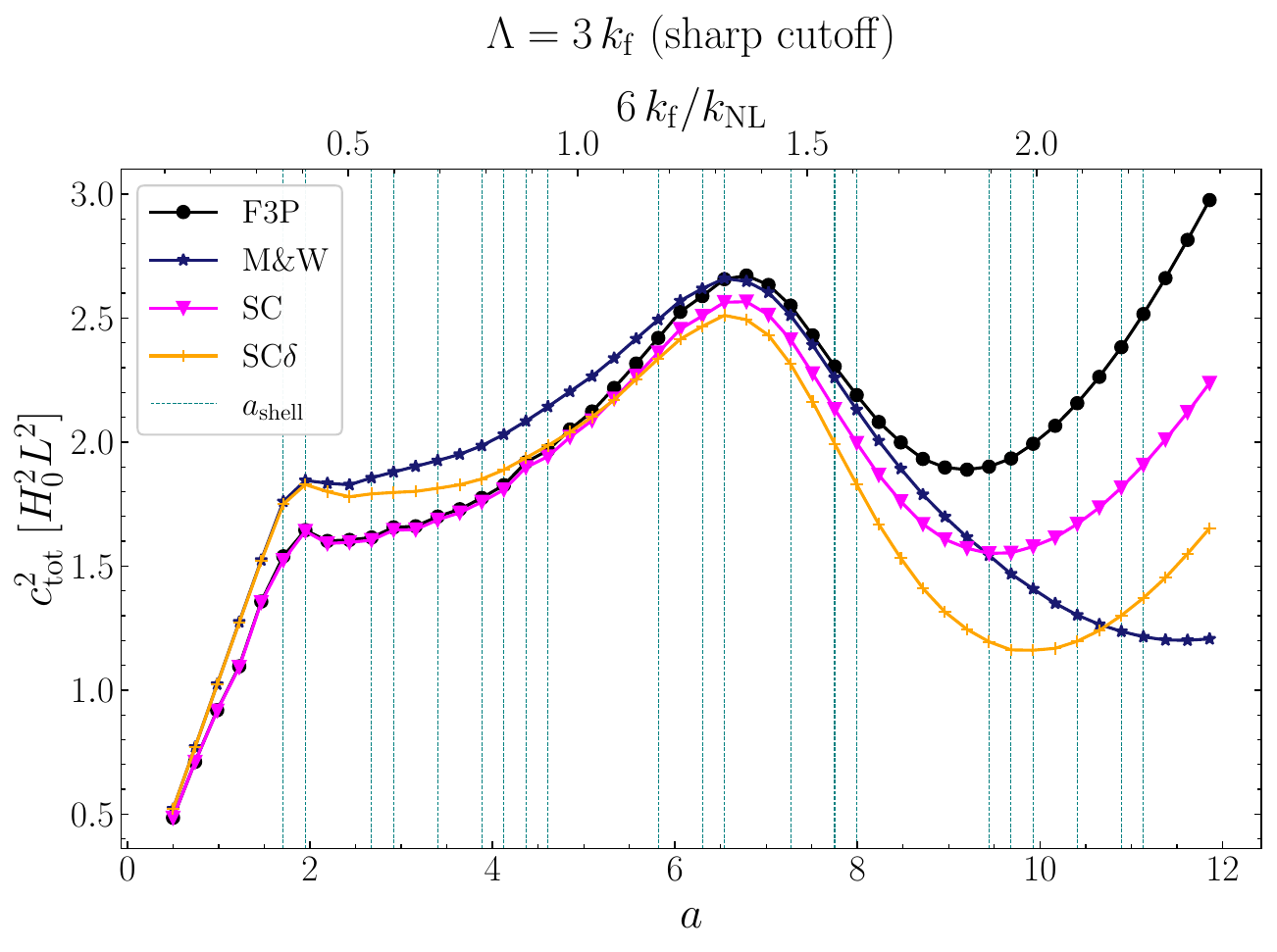}
        \caption{The long-term behaviour of $c_{\mathrm{tot}}^{2}$ for \texttt{sim\_1\_6\_15}.}
        \label{fig:ctot2_ev_long_three_modes}
    \end{figure}

We make one further test by adding an intermediate non-zero perturbation with wavenumber $6\,k_{\mathrm{f}}$ and amplitude $0.25$ to \texttt{sim\_1\_15}. We use this simulation instead of \texttt{sim\_1\_11} to increase the wavenumber separation between the non-zero perturbations so that another peak in $c_{\mathrm{tot}}^{2}$ corresponding to $6\,k_{\mathrm{f}}$ is clearly visible. The $c_{\mathrm{tot}}^{2}$ measured from this run (which we christen \texttt{sim\_1\_6\_15}) is shown in Fig. \ref{fig:ctot2_ev_long_three_modes}. As before, $c_{\mathrm{tot}}^{2} \propto a$ holds until the first shell crossing, then $c_{\mathrm{tot}}^{2}$ decreases slightly. However, the $6\,k_{\mathrm{f}}$ mode enters the non-linear regime shortly after, and we see a second peak around $a\approx 6.75$. The $x$-axis (top) shows that $k_{\mathrm{NL}} < 6$ at this epoch. After this, $c_{\mathrm{tot}}^{2}$ decreases again, until $3\,k_{\mathrm{f}}$ enters the non-linear regime (when $6\,k_{\mathrm{f}}/k_{\mathrm{NL}}$ exceeds two) as in \texttt{sim\_1\_11} leading to another increase beyond $a\approx 10$. These results seem to confirm our earlier claim that $c_{\mathrm{tot}}^{2}$ is sourced by modes entering the non-linear regime.

\section{Summary}
\label{sec:summary}
The EFTofLSS has been successful in extending the range of accurate perturbative predictions of summary statistics of the dark-matter density field to larger wavenumbers \cite[e.g.][]{baumann_2012, carrasco_2012, carrasco2014, baldauf2015bispectrum, alkhanishvili2022reach}. With extensions to include biased tracers \cite[e.g.][]{assassi2014renormalized, senatore2015bias, angulo2015statistics, mirbabayi2015biased, donath2020biased} and redshift-space distortions \cite[e.g.][]{senatore2014redshift, lewandowski2018eft}, the EFTofLSS is a valuable tool in extracting cosmological parameters from data obtained from current and next-generation surveys \cite[e.g.][]{ivanov2020cosmological, ivanov2020cosmological2, colas2020efficient, nishimichi2020blinded, chen2022new}. In this study, our motivation has been to strip away the complexity of the systems that the approach is usually applied to, and make controlled tests in the simplest possible setting. 

The EFTofLSS makes three important assumptions: one, that the physical scales are organised in a hierarchy ordered as $k \ll \Lambda \ll k_{\mathrm{NL}}$ (for cutoff scale $\Lambda$); two, that the effective stress tensor can be expanded in terms of long-wavelength fields in an expansion of the form given in Eq.~(\ref{eq:tau_expansion}); and three, that the EFT correction to a summary statistic (e.g., the power spectrum) at a given loop order regulates the UV sensitivity of the corresponding SPT term.

We specialise to an EdS universe with plane-parallel cosinusoidal initial conditions. Using five different top-down estimators, all assuming the expansion of the effective stress, we measure the full time-evolution of the EFT coefficients. For all these estimators, $c_{\mathrm{tot}}^{2}$ increases linearly with the expansion factor of the background universe until the first shell crossing, which is the scaling expected in the literature. However, after this time, we notice a clear departure from this behaviour, as $c_{\mathrm{tot}}^{2}$ decreases after $a \approx 2$ (Figure \ref{fig:ctot2_ev_sharp}).

We compute the next-to-leading-order power spectrum in SPT and in the EFTofLSS, and compare this to the power spectrum measured in our simulations (Figure \ref{fig:spt_spec}). We find that for the largest scale in our model universe, the SPT power matches the $N$-body power accurately until $a\approx3$ for our primary simulation \texttt{sim\_1\_11}. At later times, we note deviations from the $N$-body spectrum which rises to $\approx 2.2\%$ at $a=6$. Introducing a cutoff in the SPT loop integrals leads to an overprediction of the power at all times. At $a=6$, this tSPT model differs from the $N$-body spectrum by $\approx 1.5\%$. On the other hand, the EFTofLSS power spectrum, calculated via the top-down coefficients is accurate to within $\approx 0.06\%$ at all times, indicating that the EFTofLSS accurately models the feedback from the short-wavelength modes onto the large scales. Further, the top-down and bottom-up approaches are in good agreement with each other for the bare coefficient $\alpha_{c}$ (Figure \ref{fig:alpha_c_sharp}).

We also investigate the effect of separation (in wavenumber) between the initially non-zero perturbations on the EFTofLSS predictions (Figure \ref{fig:spec_comp}). As expected, the tSPT power becomes less accurate (compared to $N$-body) with smaller separation. The EFT power shows a similar trend, with the highest accuracy at largest separation. However, for all simulations, the EFT prediction is accurate to within $\approx 0.1\%$ of that of the $N$-body, showing that the EFT predictions are robust to change in separation.

We explicitly measure the stochastic fluctuations of the effective stress within our setup and find that their contribution to the matter power spectrum is much smaller than the leading-order EFT correction. Further, the power spectrum of these fluctuations cannot renormalise the $P_{22}$ term in our setup. This may be due to our initial conditions having too few degrees of freedom.

Using the measured power spectrum of \texttt{sim\_1\_11}, we calculate the renormalised leading EFT correction to the power spectrum ($\alpha_{\mathrm{ren}}$) and the associated counterterm ($\alpha_{\mathrm{ctr}}$; which cancels the UV sensitivity of $P^{\mathrm{tSPT}}_{13}$), and find that the two terms -- unsurprisingly -- do not share the same time evolution.

All of our analysis has been performed for both sharp cutoff and for Gaussian smoothings. We find differences between the estimated $\langle\tau\rangle$ from the two kernels (Figure \ref{fig:tau_fit}), which can be understood via leakage from the non-zero high-$k$ modes that is only a feature of Gaussian smoothing. However, the two cutoffs produce qualitatively similar results for the power spectrum of the overdensity field, with small differences for $a > 3$.

Finally, we assess the long-term behaviour of $c_{\mathrm{tot}}^{2}$ using a test run with the same setup as \texttt{sim\_1\_11} but extended to $a\approx 12$. We find that $c_{\mathrm{tot}}^{2}$ increases after $a=6$, likely due to the $3\,k_{\mathrm{f}}$ mode entering the non-linear regime (this is our cutoff scale, so modes bigger than $3\,k_{\mathrm{f}}$ are excluded in the sharp cutoff case). To confirm our suspicion that $c_{\mathrm{tot}}^{2}$ is sourced by modes entering the non-linear regime, we introduce an intermediate non-zero perturbation (with wavenumber $6\,k_{\mathrm{f}}$) to \texttt{sim\_1\_15}. This introduces an additional peak in the $c_{\mathrm{tot}}^{2}$ vs $a$ plot when this mode becomes non-linear, further solidifying our argument.

In summary, the EFTofLSS accurately models the feedback of the small scales onto the large-scale mode in our physical system. However, the procedure of renormalising the stochastic term does not apply in our setup, perhaps due to the simplicity of our initial conditions. Tests with a more general initial setup may provide insight into these aspects of the theory.

\acknowledgments
The authors declare no conflicts of interest or external support in the preparation of this manuscript. MK is part of the International Max Planck Research School in Astronomy and Astrophysics.

\bibliographystyle{bibstyle_jcap}
\bibliography{refs}

\appendix
\section{Results with Gaussian smoothing}\label{sec:gauss_appendix}
\begin{figure}
        \centering
        \includegraphics[scale=0.5]{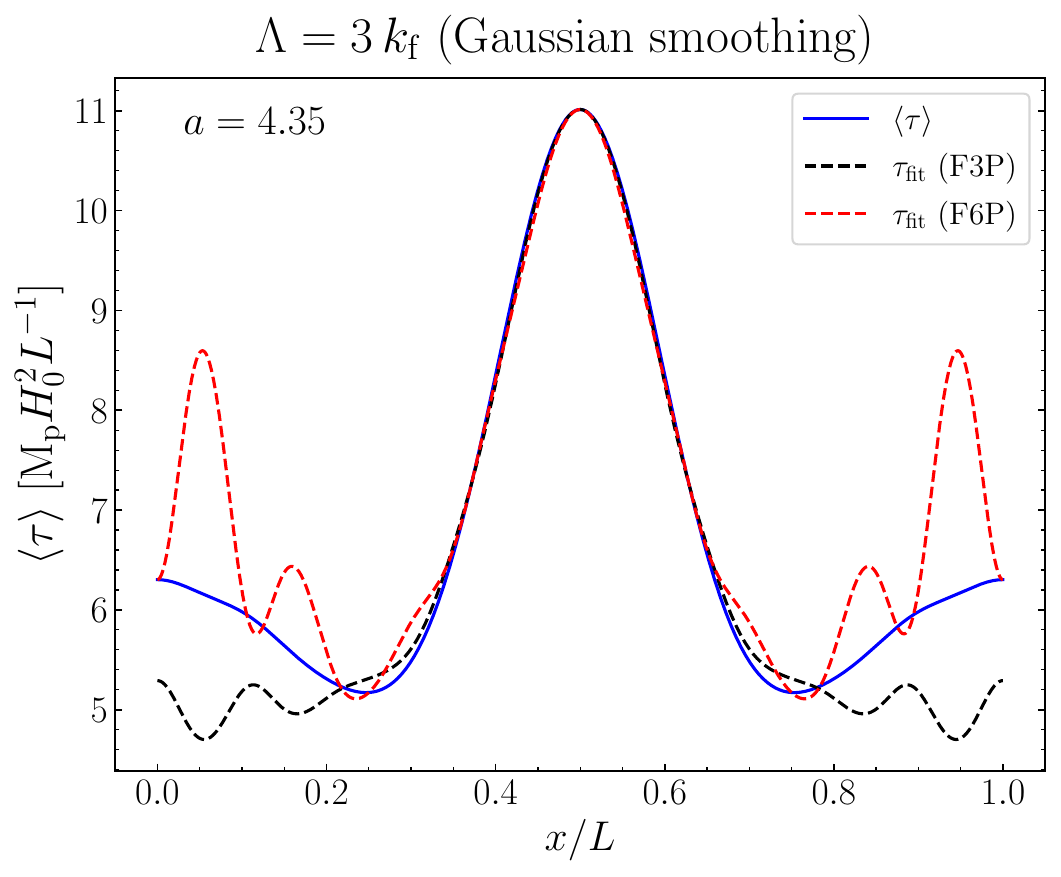}
        \caption{As in Figure~\ref{fig:tau_fit}, for $a=4.35$ with Gaussian smoothing. Note the oscillations at the outer edges of the box for the F6P fit.}
        \label{fig:tau_fit_gaussian_F6P}
    \end{figure}

In this Appendix, we show the measured EFT coefficients and the power spectrum results for Gaussian smoothing. As pointed out in Section \ref{sec:fit_to_sims}, the primary difference between the two smoothing kernels is that the Gaussian kernel allows modes with $k \gtrsim \Lambda$ to be non-zero. Thus, the smoothed $\delta$ and $\theta$ include contributions from these modes which introduce oscillatory features. These oscillations appear in the fits to $\tau$, and affect the estimation of the EFT coefficients. The effect is enhanced at second order in the expansion of $\tau$, as terms of order $\delta^{2}$ and $\theta^{2}$ are included. We show this for $a=4.35$ in Figure~\ref{fig:tau_fit_gaussian_F6P}. Note that the estimator fits the measured stress well in the central regions, but the effect of the oscillations is visible on the outskirts of the box. Due to this mis-estimation we omit the six-parameter fit (F6P) for the Gaussian smoothing results that follow, even though it is a good fit to $\tau$ for many snapshots of the simulation.

\begin{figure}
        \centering
        \includegraphics[scale=0.5]{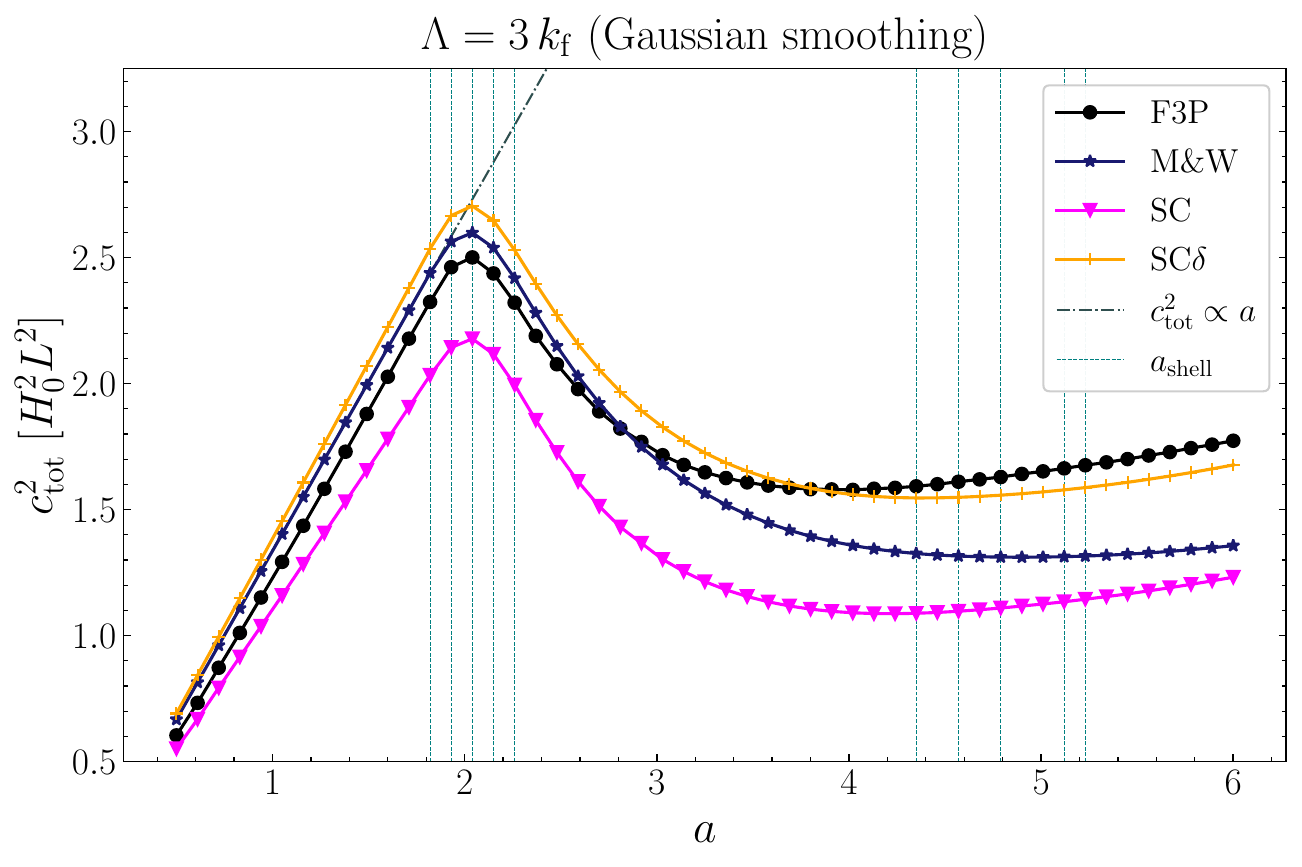}
        \caption{As in Figure~\ref{fig:ctot2_ev_sharp} but for a Gaussian filter.}
        \label{fig:ctot2_ev_gauss}
    \end{figure}

In Figure~\ref{fig:ctot2_ev_gauss}, we show the time evolution of the estimated $c_{\mathrm{tot}}^{2}$ for four top-down estimators, analogous to Figure \ref{fig:ctot2_ev_sharp}. We notice the same general behaviour as in the sharp case, viz. that $c_{\mathrm{tot}}^{2}$ increases linearly until the first shell crossing, reaches a maximum around $a\approx2$, then decreases with increasing $a$. However, there are two main differences from the sharp case: one, the estimators have a larger variation among them, and two, the value of $c_{\mathrm{tot}}^{2}$ plateaus after an initial decrease post the first shell crossing, then shows a slight increase at very late times.

\begin{figure}
        \centering
        \includegraphics[scale=0.5]{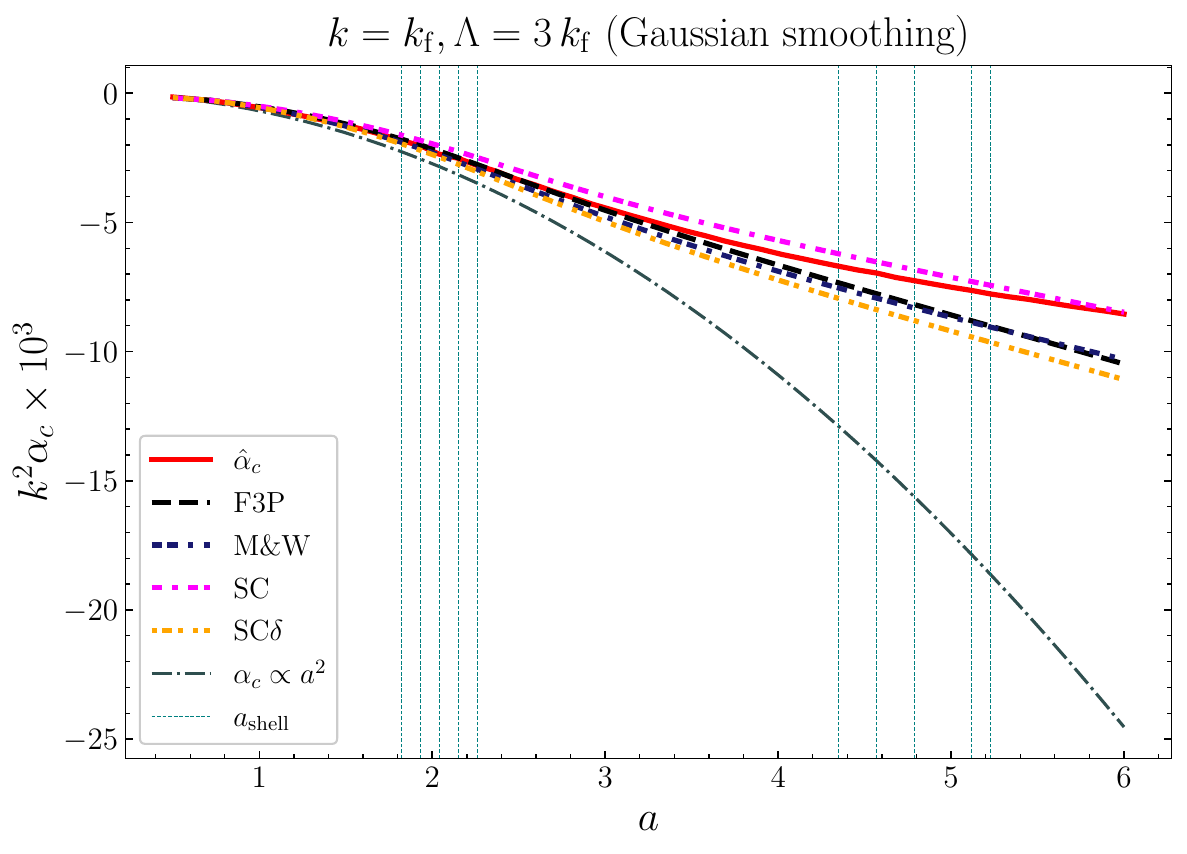}
        \caption{As in Figure~\ref{fig:alpha_c_sharp} but for a Gaussian filter.}
        \label{fig:alpha_c_gauss}
    \end{figure}

The time evolution of $\alpha_{c}$ is shown in Figure \ref{fig:alpha_c_gauss} for the bottom-up (dashed lines) and top-down (solid line) estimators. As in the corresponding sharp-cutoff plot (Figure \ref{fig:alpha_c_sharp}), the top-down estimators are in rough agreement with each other and with the bottom-up estimator. However, there is larger variation among them compared to the sharp case, consistent with the observed behaviour of $c_{\mathrm{tot}}^{2}$.

\begin{figure}
        \centering
        \includegraphics[scale=0.55]{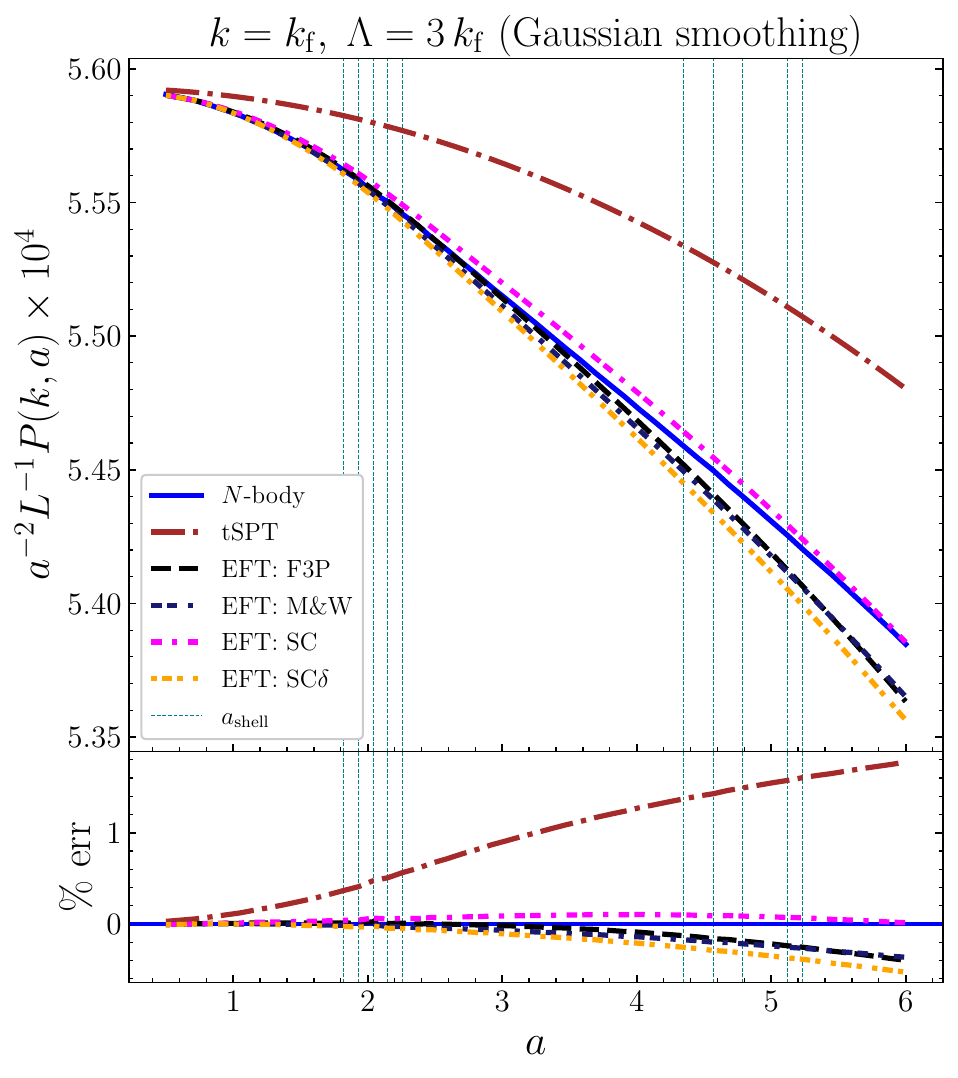}
        \caption{As in Figure~\ref{fig:spt_spec} but for a Gaussian filter.}
        \label{fig:eft_spectra_k1_L3_gauss}
    \end{figure}

Finally, in Figure \ref{fig:eft_spectra_k1_L3_gauss}, we plot the non-linear power in the $k=k_\mathrm{f}$ mode in the different theories for the Gaussian case, in analogy with the right panel of Figure \ref{fig:spt_spec}. The agreement between the top-down and bottom-up estimators is good until $a\approx3$, after which most of our estimators over-correct the tSPT-4 power. The SC estimator remains accurate to within $\approx 0.1\%$ throughout the simulation run.
Overall, the results from Gaussian smoothing complement the results from sharp case in the main study. Although the match between the top-down and the bottom-up estimates is slightly worse than in the sharp case (especially at late times), the choice of the smoothing kernel does not influence the results strongly.

\end{document}